\documentclass[aps,prx,twocolumn,superscriptaddress,longbibliography]{revtex4-2}
\usepackage[pages=all, color=black, position={current page.south}, placement=bottom, scale=1, opacity=1, vshift=5mm]{background}
\SetBgContents{
	\tt 
}      

\usepackage{amsmath}
\usepackage{amsthm}
\usepackage{amssymb}
\usepackage{times}
\usepackage{braket}
\usepackage{ulem}

\usepackage{tikz}
\usetikzlibrary{quantikz2} 

\usepackage[utf8]{inputenc}
\usepackage{hyperref}

\usepackage{graphicx, color}
\graphicspath{{fig/}}

\usepackage{algorithm, algpseudocode} 
\usepackage{mathrsfs} 
\usepackage{lipsum}

\begin{document}

\title{Photon distillation schemes with reduced resource costs based on multiphoton Fourier interference}

\author{F.~H.~B.~Somhorst}
\affiliation{MESA+ Institute for Nanotechnology, University of Twente, P.~O.~box 217, 7500 AE Enschede, The Netherlands} 
\author{B.~K.~Sauër}
\affiliation{MESA+ Institute for Nanotechnology, University of Twente, P.~O.~box 217, 7500 AE Enschede, The Netherlands} 
\author{S.~N.~van den Hoven}
\affiliation{MESA+ Institute for Nanotechnology, University of Twente, P.~O.~box 217, 7500 AE Enschede, The Netherlands} 
\author{J.~J.~Renema}
\email{j.j.renema@utwente.nl}
\affiliation{MESA+ Institute for Nanotechnology, University of Twente, P.~O.~box 217, 7500 AE Enschede, The Netherlands} 


\begin{abstract}
    We present a scalable scheme to achieve photon distillation, i.e. the preparation of a single photon with reduced indistinguishability error out of multiple imperfect photons, based on multiphoton interference in Fourier matrices. Our scheme achieves arbitrary error reduction in a single step, removing the need to concatenate multiple rounds of the protocol. Our scheme has resource costs linear in the error reduction, and a constant probability of success for arbitrary error reductions. We find that the scheme both raises the threshold of error correcting codes in photonic quantum computers, as well as reduces the resources required to construct a photonic quantum computer, making it highly promising for implementation in such devices. 
\end{abstract}

\maketitle

\section{Introduction}
\label{sec:intro}

Since the introduction of the Knill-Laflamme-Milburn  scheme, linear optics stands out as a promising hardware platform for universal, fault-tolerant photonic quantum computing \cite{knill2001scheme}, due to the long coherence time of photons and the high-precision control available over optical quantum states. In linear optics, the allowable operations are restricted to measurement, linear optical circuits, i.e. optical devices which act linearly on the electric field amplitude, and classical reconfiguration of such circuits based on earlier measurement outcomes. Modern efforts center on fusion-based quantum computing (FBQC) \cite{browne2005resource,rudolph2017optimistic,bartolucci2023fusion}, a computational architecture where multiple fixed-size entangled states - resource states - are entangled by fusion measurements. Recent experimental progress, for example in experimentally generating noisy three-qubit entangled states \cite{chen2023heralded, cao2023photonic, maring2023general}, shows the potential of this protocol. 

The key resource for all linear-optical quantum computing schemes is photon indistinguishability \cite{mandel1991coherence,shchesnovich2014sufficient,lal2022indistinguishable}, i.e. the degree to which the photons in the quantum computer are identical in the degrees of freedom of those photons which are not being used to encode information (e.g. polarization, time, and frequency \cite{flamini2018photonic}). Photonic indistinguishability is essential because while linearity forbids an explicit photon-photon interaction, it allows for multi-particle quantum interference, which is sufficient for universal quantum computation when combined with measurements and fast feedback. The archetypal example of such interference is the Hong-Ou-Mandel effect \cite{hong1987measurement}. Since photon indistinguishability  governs the degree to which multi-particle interference occurs, it is the crucial resource for photonic quantum computing in its modern implementations \cite{kok2007linear,o2007optical,varnava2008good,slussarenko2019photonic}, and imperfect indistinguishability gives rise to logical errors in the computation.

\begin{figure}
    \centering
     \includegraphics[width=8.6cm, height=20cm, keepaspectratio,]{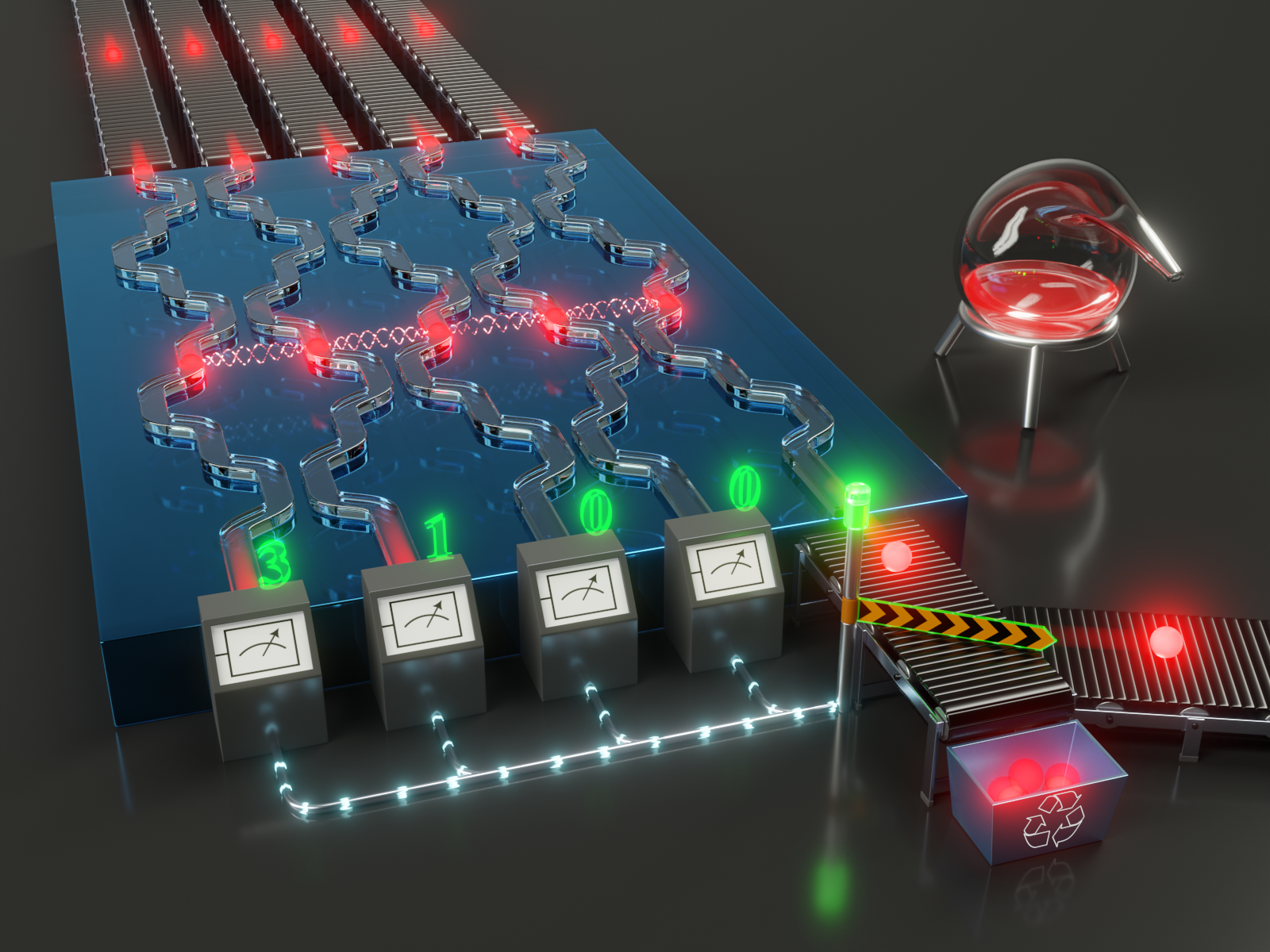}
    \caption{\textbf{Artist's impression of photon distillation.} Partially distinguishable photons emitted by imperfect sources are interfered by a photonic distillation circuit. A given set of partial measurements heralds a single photon state with enhanced indistinguishability. After a successful distillation attempt, the distilled photon, with reduced indistinguishability error, is ready for processing in photonic quantum computing.}
    \label{fig:fig1_schem}
\end{figure}

To improve the degree of indistinguishability of photons, photonic distillation schemes 
\cite{sparrow2017quantum,marshall2022distillation} have been proposed and experimentally demonstrated \cite{faurby2024purifying}. Photon distillation (see Fig. \ref{fig:fig1_schem}) is a nondeterministic method for heralding photons with a reduced indistinguishability error. These schemes trade  numbers for quality, using interference and measurement to exchange multiple weakly indistinguishable photons for one photon with reduced distinguishability error, analogous to magic state distillation \cite{bravyi2005universal}.
In contrast to optical filtering, photon distillation does not require prior knowledge about the target wave function of the indistinguishable photons, and filters in all internal degrees of freedom simultaneously. However, existing photon distillation protocols  \cite{sparrow2017quantum, marshall2022distillation} suffer from substantial resource costs due to the fact that each round of the protocol only moderately reduces the indistinguishability error, requiring concatenation to achieve higher error reduction. Since each round fails with a constant probability, this requires multiplexing \cite{migdall2002tailoring} and intermediate optical feed-forward, which results in resource requirements which render the protocol infeasible in practice. 

In this work, we present a photon distillation scheme with significantly reduced resource costs, rendering the protocol feasible for implementation in photonic quantum computers. Our photon distillation scheme requires no additional technological developments beyond those already needed for fault-tolerant photonic quantum computing architectures. In practice, the resource costs of our quantum error mitigation protocol is expected to be only a small fraction of the total resource costs of the architecture \cite{faurby2024purifying}. Our scheme is based on multiphoton interference in Fourier matrices. In contrast to the state of the art \cite{marshall2022distillation}, our scheme can achieve arbitrary reduction of errors in a single round of the protocol, thereby eliminating the need for concatenation of multiple distillation steps or intermediate fast feedback. Up to a conjecture in the theory of matrix permanents, we show that the resource costs as expressed in the overall number of photons required for the protocol are linear in the required error reduction,  and the success probability of the protocol tends to a constant as more photons are used. These two facts render our protocol highly scaleable: to achieve an asymptotic reduction in indistinguishability error by at least a factor of 100 required 59,000 photons with the state-of-the-art scheme, whereas our new scheme only requires 400 photons. We show that our scheme is asymptotically optimal in the degree of error reduction per photon consumed.

These results are based on the study of multiphoton interference in Fourier matrices, using simulation techniques originally designed for boson sampling \cite{shchesnovich2014sufficient, shchesnovich2015partial,tichy2015sampling, renema2018efficient, moylett2019classically}. We follow up on the observation that the scheme proposed in \cite{marshall2022distillation}, originally found by numerical optimization, in fact implements a Fourier transform on 3 photons in 3 optical modes. The key difference between interference in small systems (with 3 or 4 photons) and larger system is that in larger systems, multiple herald outcomes are valid for distillation, rendering large distillation schemes much more efficient than expected based on extrapolating the smaller schemes. 

Furthermore, we study the application of our protocol in large-scale, fault-tolerant, universal photonic quantum computers, specifically in comparison to quantum error correction (QEC). We find evidence that photon distillation has a higher threshold for distinguishability errors than the surface code. Moreover, we find that a regime exists below the error threshold of the surface code where a combination of photon distillation and QEC is more resource-efficient than only QEC. We discuss how our scheme could operate in practice outside the limit of small errors. These results suggest that an implementation of our photon distillation protocol is perfectly tailored for seamless integration into large-scale photonic quantum computers. 

\textit{Note added.} - Related work was simultaneously released \cite{saied2024general}.

\section{Interference of partially distinguishable photons}

We begin by introducing a model for partially distinguishable photons. The \textit{orthogonal bad-bit model} describes single photon states produced by imperfect sources \cite{sparrow2017quantum}. The internal error modes do not contribute to multiphoton interference because of their mutual orthogonality, thus allowing us to expand the imperfect single photon states $\rho(\epsilon)$ into equivalent sums of fully indistinguishable $\rho_\text{i} = \ket{\psi_0}\bra{\psi_0}$ and fully distinguishable $\rho_\text{d}$ photons \cite{moylett2019classically}
\begin{equation}
    \rho(\epsilon) = (1- \epsilon)\rho_\text{i} + \epsilon \rho_\text{d}, 
    \label{eq:SingleMoylett}
\end{equation}
where $\epsilon$ is the parameter used in this work for the partial distinguishability error \footnote{The \textit{orthogonal bad-bit model} is equivalent to the \textit{partial distinguishability model} where $x := 1 - \epsilon$  
\cite{renema2018efficient}. The assumption of identical mutual partial distinguishability is demonstrated to be appropriate in multiphoton experiments with imperfect sources of sufficient quality \cite{renema2021sample}.}. A combination of an optical transformation plus partial measurement which projects onto a single photon state $\rho(\epsilon^\prime)$ results in photon distillation when $\epsilon^\prime < \epsilon$ \footnote{We note that for a distilled photon, $\rho_{\text{d}}$ is no longer a pure state but instead a maximally mixed state over the error modes of the initial resource photons. However, the uniqueness of these error modes is preserved in the mixed error state of the distilled photon. Consequently, the precise nature of $\rho_{\text{d}}$ is irrelevant; for subsequent multi-photon interference, it is sufficient to treat $\rho_{\text{d}}$ as equivalent to a pure, fully distinguishable photon state.}.  \\

In linear quantum optics, optical transformations $U$ are implemented as a linear transformation acting on the bosonic modal creation operators $(\hat{a}_1^\dagger, \dots, \hat{a}_N^\dagger)^T \mapsto U(\hat{a}_1^\dagger, \dots, \hat{a}_N^\dagger)^T$, with $U$ a unitary matrix. In state space representation, the imperfect multiphoton state transforms to
\begin{equation}
    \rho(\epsilon)^{\otimes N} \mapsto \sigma(\epsilon) := \mathcal{U}(U) \rho(\epsilon)^{\otimes N}\mathcal{U}(U)^\dagger. 
\end{equation}

Where $\mathcal{U}(U)$ is the state space transformation induced by the modal transformation $U$  \cite{arvind1995real}. For given partial measurements in photon number basis, we project the remaining mode $N$ into the state
\begin{equation}
        \rho(\epsilon^\prime) = {\mathcal{N}}^{-1}\langle n_1, \dots, n_{N-1}| \sigma(\epsilon) | n_1, \dots, n_{N-1} \rangle,
        \label{eq:distilledrho}
\end{equation}
with normalization $\mathcal{N} = \text{tr}[|n_1, \dots, n_{N-1}\rangle\langle n_1, \dots, n_{N-1}|\sigma(\epsilon)]$ . Here we have assumed (as is the case in experiments) that the internal quantum state of the photons in modes $1, \dots, N-1$ cannot be resolved. Because $U$ conserves photon number, the state in the remaining mode is projected into a definite photon number. Particularly, if the condition $\sum_{i=1}^{N-1} n_i = N-1$ is obeyed, the state is projected onto a single photon, with an internal state $\rho(\epsilon')$ which can be computed from Eq. \ref{eq:distilledrho}.

Finally, we observe that for sufficiently low partial distinguishability error $\epsilon$, the input state can be approximated as
\begin{equation}
	\rho(\epsilon)^{\otimes N} = (1 – N \epsilon) {\rho_\text{i}}^{\otimes N} +  \epsilon \sum_{k=1}^{N}{\rho_\text{i}}^{\otimes k-1}\otimes{\rho_\text{d}}\otimes{\rho_\text{i}}^{\otimes N-k} + \mathcal{O}(\epsilon^2).
    \label{eq:MultiMoylett}
\end{equation}
In App. \ref{appendix:firstordererror} we derive that this first-order error approximation is valid for photon distillation if $\delta_N(\epsilon) \ll \epsilon$, where $\delta_N(\epsilon) := 1 – (1-\epsilon)^N – N\epsilon(1-\epsilon)^{N-1}$ is the probability of higher-order errors in the input state.\\

Next, we recall some facts about Fourier transformations in linear optics. For a Fourier transformation, the matrix elements of $U$ read:
\begin{equation}
    U_{jk} := \frac{1}{\sqrt{N}} e^{i2\pi (j-1)(k-1)/N}.
    \label{eq:Fourierdef}
\end{equation} 

Such transformations have been studied extensively \cite{tichy2010zero, tichy2014stringent, carolan2015universal, crespi2016suppression, stanisic2018discriminating,dittel2018totally, brod2019witnessing, somhorst2023quantum} in quantum optics, since their high degree of symmetry offers various attractive properties. In particular, when a single, fully indistinguishable photon is present in each input mode, the output distribution of a Fourier matrix obeys the \textit{zero-transmission law}, where all outcomes which have 
\begin{equation}
	{\sum_{k = 1}^{N}m_k \text{ mod } N} \neq 0,
    \label{eq:nonforbiddenstates}
\end{equation}  
are never observed \cite{tichy2010zero}, where $m = (m_1, \dots, m_N)$ is the equivalent ordered mode-assignment list corresponding to a particular measurement $|n_1, \dots, n_{N}\rangle \langle n_1, \dots, n_{N}|$ \footnote{For example, the mode-assignment list $(1,2,3,4,5)$ corresponds to the equivalent non-collision state $\ket{11111}$.}. Such outcomes are referred to as \textit{forbidden outcomes}, and we will refer to the complementary set of these outcomes as \textit{allowed outcomes} \footnote{However, a classification as \textit{allowed outcome} is a necessary but not sufficient condition for a classification as \textit{non-suppressed outcome}. Nevertheless, most allowed outcomes constitute non-suppressed outcomes \cite{tichy2010zero}.}. The fact that occurrence of these outcomes can be predicted efficiently despite the fact that computing the probability of a particular outcome of a linear optical circuit is in general computationally hard, combined with the fact that distinguishable photons do not obey the zero-transmission law, makes Fourier matrices an attractive probe for photon distinguishability in experiments. Moreover, straightforward numerical optimization of the task of estimating photonic indistinguishability results in either Fourier matrices or combinations of Fourier matrices being the optimal choice of matrix settings, depending on the precise specifications of the task \cite{stanisic2018discriminating}.

\section{Fourier transform-based distillation}
\label{sec:FTBP}

In the previous section, we established the framework for imperfect photon interference. To establish intuition for the problem of photon distillation in Fourier matrices, we first discuss the relevant metrics for distillation circuits. We illustrate these metrics by extensively studying the $N=5$ photon Fourier transform with a single photon at each input mode, in both error reduction and heralding probability. A detailed description of our numerical simulation methods is presented in App. \ref{appendix:simulationmethods}.

The resource cost $C$ of photon distillation schemes is given by the number of photons of a given degree of indistinguishability that is on average required to obtain one photon of a given improved indistinguishability, or equivalently, the degree of improvement that can on average be achieved with a fixed number of photons. This resource cost, in turn, is determined by two factors: the \textit{error reduction} $R$, i.e. the degree to which a single successful application of the scheme reduces the indistinguishability error on each photon, and the \textit{heralding probability} $p$, i.e. the probability with which the scheme succeeds in the first place. In the limit of low errors, the error reduction reduces to a multiplicative factor, i.e. $\epsilon' = \epsilon / R$, where $\epsilon$ and $\epsilon'$ are the degree of error before and after the protocol, respectively. Simultaneously, the associated resource cost for a single distillation step is $C = N/p$. 

The current state-of-the-art Fourier transform-based photon distillation protocol is presented in \cite{marshall2022distillation}, with a degree of error reduction of $R = 3$ and a heralding success probability $p = \frac{1}{3}$. Since the available error reduction is limited, further error reduction is available by concatenations, i.e. using the input from one distillation step as the input for the next. However, concatenation has a resource cost quadratic in the overall error reduction required.

\begin{figure}[h!]
    \centering
    \includegraphics[width=8.6cm, height=20cm, keepaspectratio,]{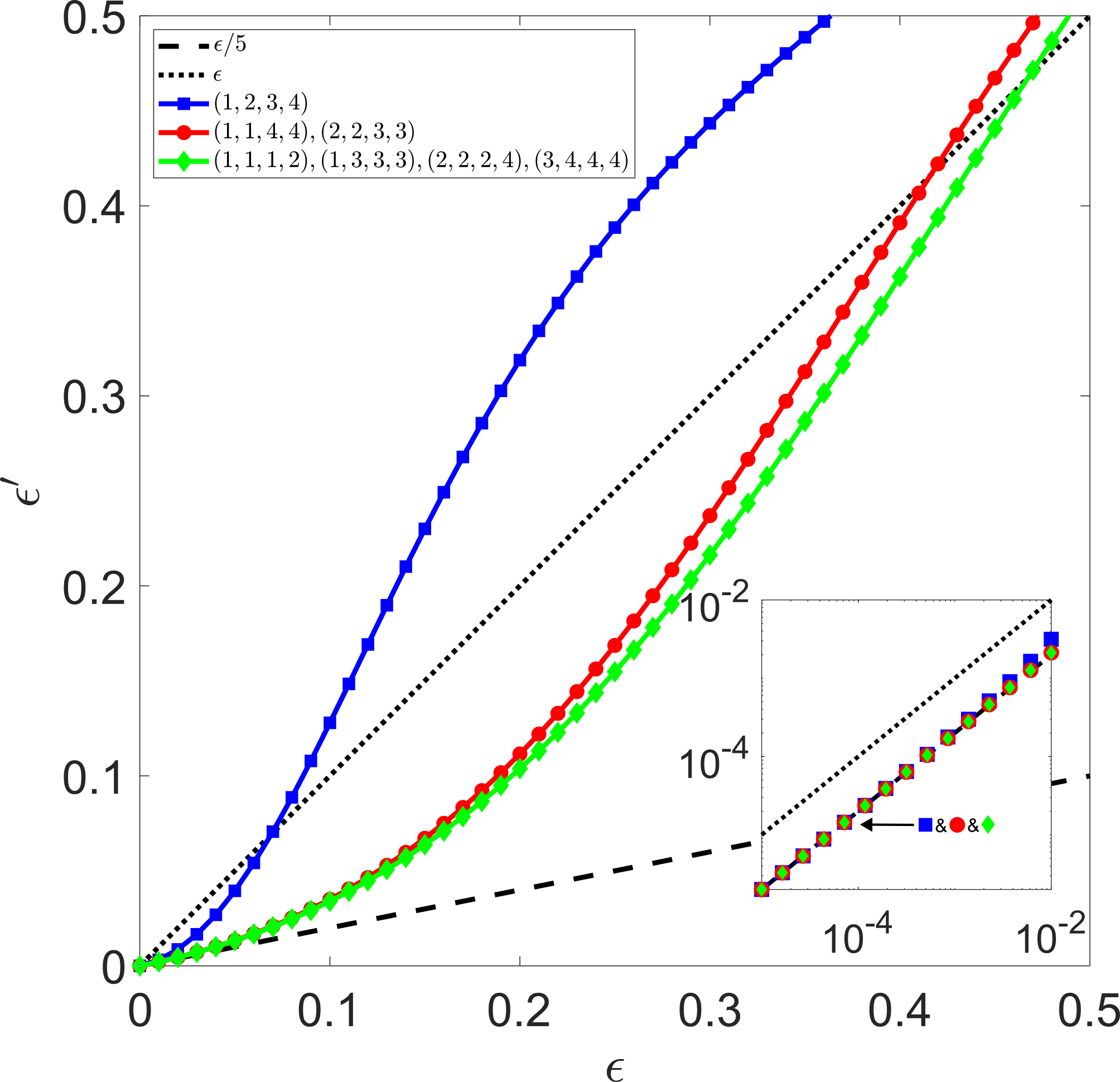}
    \caption{\textbf{Error reduction for the 5-photon Fourier transform-based distillation scheme.} One photon with partial distinguishability error $\epsilon^\prime$ is distilled out of $N = 5$ photons with initial error $\epsilon$.  Here, the legend indicates (ordered) herald measurement mode-assignment lists $(m_1, m_2, m_3, m_4)$ that specify how the remaining 4 photons are distributed among the first 4 modes, where the distilled photon is in mode 5. All other measurements result in a failure of the distillation scheme. The distillation process only works when $\epsilon$ is below a herald measurement-dependent threshold. \textit{Inset} Zoom-in to the low error regime, on a double-logarithmic scale. In this regime, we observe identical linear behaviour for all distillation patterns.}
    \label{fig:fig3_resulterror}
\end{figure}

Fig. \ref{fig:fig3_resulterror} shows the error reduction as a function of input error for the $N = 5$ photon Fourier transform-based distillation circuit, for different partial measurement outcomes on  modes 1 through 4. Surprisingly, there is more than one valid herald measurement for successful photon distillation. We observe three distinct families of successful heralding patterns. 

We observe a correspondence between allowed outcomes following Eq. \ref{eq:nonforbiddenstates} and herald patterns which correspond to photon distillation: every herald pattern which enables photon distillation corresponds to an allowed outcome. In the limit of small initial error (see inset of Fig. \ref{fig:fig3_resulterror}), we find an error reduction $\epsilon^\prime = \frac{\epsilon}{5} + \mathcal{O}(\epsilon^2)$ for all distinct herald measurements, accurate to the numerical precision of our computations. 

\begin{figure}[h!]
    \centering
    \includegraphics[width=8.6cm, height=20cm, keepaspectratio,]{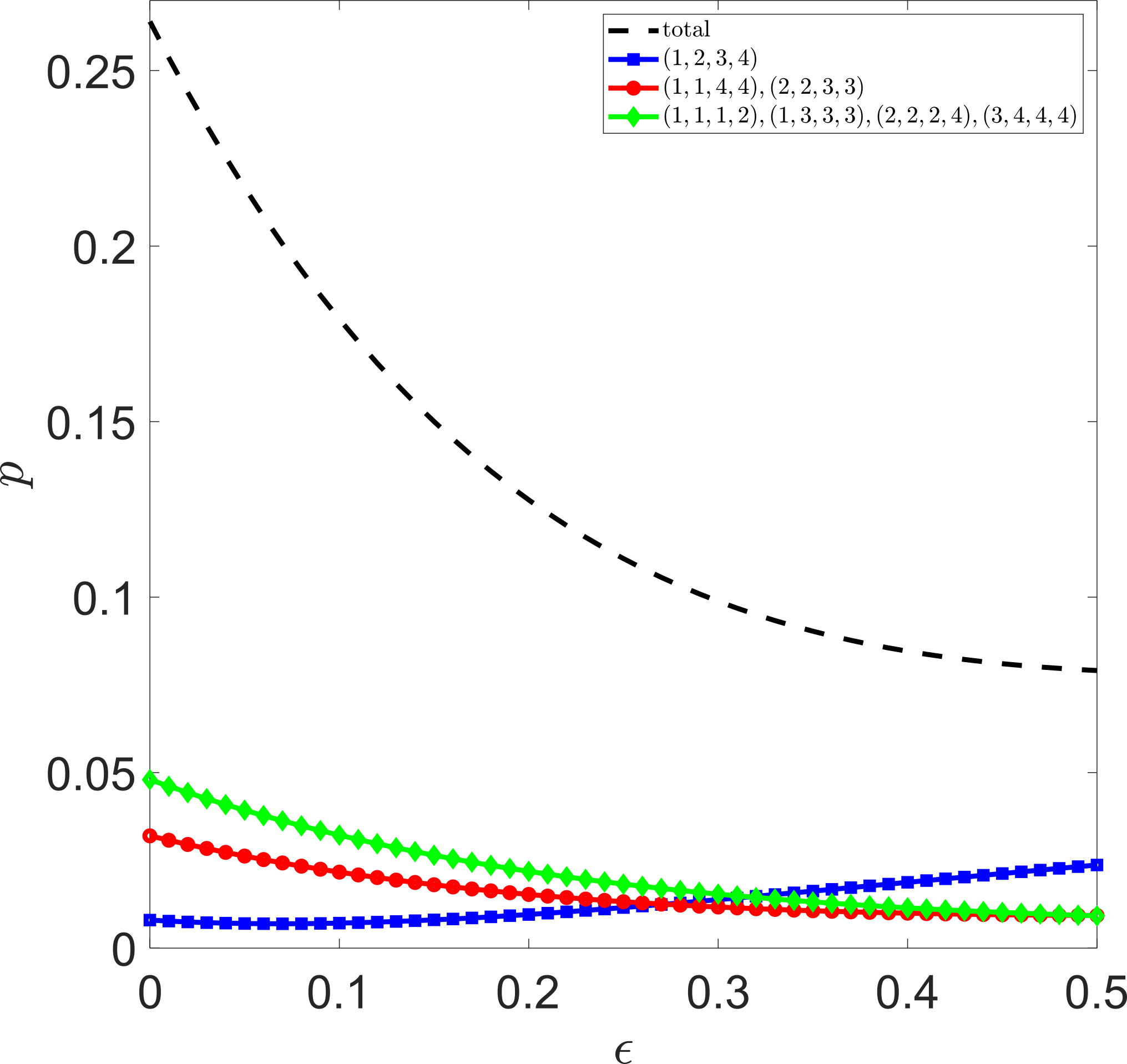}
    \caption{\textbf{Herald probability for the 5-photon Fourier transform-based distillation scheme.} This figure shows the herald probability $p$ as a function of partial distinguishability error $\epsilon$. The legend indicates (ordered) herald measurement mode-assignment lists $(m_1, m_2, m_3, m_4)$ that specify how the remaining 4 photons are distributed among the first 4 modes, while the distilled photon is in mode 5. The dashed line plots the total herald probability, which is larger than 1/4 in the low error regime. }
    \label{fig:resultheraldprob}
\end{figure}

Fig. \ref{fig:resultheraldprob} shows the heralding probabilities associated with each family of distillation outcomes. For asymptotically low initial error, we find a total herald probability of $p = \frac{33}{125} + \mathcal{O}(\epsilon)$, divided over many heralding outcomes. This breaks the previously conjectured $p = \frac{1}{N} + \mathcal{O}(\epsilon)$ scaling \cite{marshall2022distillation} coming from a single outcome. We strengthen our observations at $N = 5$ by numerical simulations up to $N = 10$. We find error scaling of $\epsilon^\prime = \frac{\epsilon}{N} + \mathcal{O}(\epsilon^2)$, while simultaneously a total herald success probability of $p \geq \frac{1}{4} + \mathcal{O}(\epsilon)$ is observed \cite{dataSet}. A phenomenon which occurs at higher $N$ which is not observed at $N = 5$ is that there are allowed outcomes which nonetheless have zero output probability - a consequence of the fact that Eq. \ref{eq:nonforbiddenstates} is a necessary but not sufficient condition. Ref. \cite{saied2024general} explicitly demonstrates that for $N$ not equal to prime powers, suppressed allowed outcomes still persist, whereas for $N$ equal to prime powers, Eq. \ref{eq:nonforbiddenstates}  provides a necessary and sufficient condition.

\section{Theoretical analysis}
\label{section:explanations}
We now concentrate on studying the limit of low partial distinguishability error. 
The following three key facts, valid in the limit of low error, emerge from our numerical study: 
\begin{itemize}
    \item The error reduction tends to $\epsilon^\prime = \epsilon / N$ for valid herald patterns.
    \item Every allowed outcome which has precisely 1 photon in the desired optical mode constitutes a valid partial measurement for distillation. 
    \item The total herald success probability tends to $\frac{1}{4}$ in the limit of large $N$. 
\end{itemize}

In this section, we will provide our evidence for these assertions. We will prove the first two claims up to a  technical conjecture in the theory of permanents of Fourier matrices, and give numerical and heuristic evidence for the third. In the next section, we will discuss the implications of these results for the design of distillation schemes in photonic quantum computers.

We will first focus on the error reduction. In the orthogonal bad-bit model, we have a classical mixture of states containing various error contributions, and the photon distillation circuit transforms this state to:

\begin{equation}
    \begin{split}
        \sigma(\epsilon) &= (1-N \epsilon) \mathcal{U} {\rho_\text{i}}^{\otimes N} \mathcal{U}^\dagger \\
        &+ \epsilon \sum_{k=1}^{N} \mathcal{U} \rho_\text{k} \mathcal{U}^\dagger + \mathcal{O}(\epsilon^2),
    \end{split}
    \label{eq:sigma_expanded}
\end{equation}
where we used $\mathcal{U} := \mathcal{U}(U)$ and $\rho_{\text{k}} := {\rho_\text{i}}^{\otimes k-1}\otimes{\rho_\text{d}}\otimes{\rho_\text{i}}^{\otimes N-k}$ for convenience. The probability of our state containing zero errors after a measurement $M$ is given by: 
\begin{equation}
    \begin{split}
        &P({\rho_\text{i}}^{\otimes N} | M) := \\
        &(1-N \epsilon) \langle n_1, \dots, n_{N-1} 1| \mathcal{U} {\rho_\text{i}}^{\otimes N} \mathcal{U}^\dagger | n_1, \dots, n_{N-1} 1 \rangle,
    \end{split}
    \label{eq:defPosterior}
\end{equation}
which signifies that there is 1 photon in mode $N$, as required for photon distillation. Since this distilled photon originates from the zero-error state, it is, by definition, projected into the indistinguishable photon state: $P(\rho_{\text{i}} | \rho_{\text{i}}^{\otimes N}) = 1$. Similarly, the probability of our state containing one error in input mode $k$ after the same measurement is given by:
\begin{equation}
        P(\rho_{\text{k}} | M) := 
        \epsilon \langle n_1, \dots, n_{N-1} 1| \mathcal{U} \rho_{\text{k}} \mathcal{U}^\dagger | n_1, \dots, n_{N-1} 1 \rangle.
    \label{eq:defPosteriorError}
\end{equation}
As the photon distillation transform is symmetric, there is no bias towards a specific output mode for the distinguishable photon. Consequently, there is a $1/N$ probability that the distilled photon will be projected into an error mode, implying that the probability of it being projected into the indistinguishable photon state is $P(\rho_{\text{i}} |\rho_{\text{k}}) = (N-1)/N$. 

The error reduction is most easily understood in a Bayesian context: the posterior probability of our state containing zero errors after a measurement $M$ is given by:
\begin{equation}
    P({\rho_\text{i}}^{\otimes N} | M) = \frac{q_\text{i}}{q(\epsilon)} \cdot P({\rho_\text{i}}^{\otimes N}),
    \label{eq:BayesianIndist}
\end{equation}
where $q(\epsilon)$ is the probability of obtaining that measurement outcome as a function of the error probability, and $q_i$ is the probability of obtaining that measurement outcome for fully indistinguishable photons. Similarly, the probability of our state containing one error after a measurement $M$ is given by:
\begin{equation}
    \sum_{k=1}^{N} P(\rho_{\text{k}}| M) = \frac{\lambda}{q(\epsilon)} \cdot \sum_{k=1}^{N} P( \rho_{\text{k}} ),
    \label{eq:BayesianDist}
\end{equation}
where $\lambda$ is the probability of obtaining the same measurement outcome given that one of the photons is distinguishable. This implies that if we observe a partial measurement outcome which is more likely to occur when the state contains zero errors, i.e. $q_{\text{i}} > \lambda$, then we increase the posterior probability that the parts of our state which remain correspond to the part that contains zero errors. We are therefore looking for partial measurements where the probability of observing that outcome, $q(\epsilon)$ is a strongly decreasing function of the error probability. Finally, the new error is defined by the conditional probability of observing an indistinguishable photon given the distilled photon state, $P(\rho_{\text{i}} | \rho(\epsilon^\prime)) := 1 - \epsilon^\prime$, which is found by evaluating:
\begin{equation}
    \begin{split}
    P(\rho_{\text{i}} | \rho(\epsilon^\prime)) &= P({\rho_\text{i}}^{\otimes N} | M) \cdot P(\rho_{\text{i}} | \rho_{\text{i}}^{\otimes N}) \\
    &+ \sum_{k=1}^{N} P(\rho_{\text{k}}| M)  \cdot P(\rho_{\text{i}} |\rho_{\text{k}}). 
    \end{split}
    \label{eq:BayesNewError}
\end{equation}

To compute the probability of a measurement outcome as a function of the probability of an error, we use the formalism of Refs. \cite{tichy2015sampling,renema2018efficient}. In this formalism, multiphoton interference is split up into terms corresponding to quantum interference of $j$ photons and classical transmission of the remaining $N-j$ photons, summed over all partitions into interfering and non-interfering photons. The probability of a single given herald measurement outcome is then given by 

\begin{equation}
\label{eq:StartIndist}
q(\epsilon) := \sum_{j=0}^{N} c_j (1-\epsilon)^j,
\end{equation}
where $c_j$ sums up the contributions from all $j$-photon interference contributions \cite{renema2018efficient}.

For small partial distinguishability error, we approximate Eq. \ref{eq:StartIndist} as:
\begin{equation}
    q(\epsilon) = \sum_{j=0}^{N} c_j - \epsilon \sum_{j=0}^{N}c_j j + \mathcal{O}(\epsilon^2),
    \label{eq:TwoStep}
\end{equation}
where $q_\text{i} = \sum_{j=0}^{N} c_j$ is the measurement probability for fully indistinguishable photons. 

We now introduce our conjecture: for Fourier matrices fed with single photons, we find that the identity 
\begin{equation}
    \sum_{j=0}^{N}c_j j = (N-1) \sum_{j=0}^{N}c_j.
    \label{eq:conjecture}
\end{equation}
holds. We have numerically tested this conjecture for $N=3$ through $N=10$ (App. \ref{app:conjecture}, Tab. \ref{tab:summary_numerical_study}) and find that it holds to within the numerical precision of our calculations. 

We use Eq. \ref{eq:conjecture} to write 
\begin{equation}
    q(\epsilon) = q_\text{i} - \epsilon (N-1) q_\text{i} + \mathcal{O}(\epsilon^2).
    \label{eq:ImplicationConjecture}
\end{equation} 
Combining this with (Eq. \ref{eq:MultiMoylett})
\begin{equation}
\label{eq:FromModel}
    q(\epsilon) = q_\text{i}(1-N \epsilon) + \lambda N \epsilon + \mathcal{O}(\epsilon^2),
    \end{equation}
we find $\lambda = q_\text{i}$/$N$. Following a similar approach as Ref. \cite{marshall2022distillation} by using Eq. \ref{eq:BayesNewError}, we find for the new error $\epsilon^\prime$
\begin{equation}
    1 - \epsilon^\prime = \frac{(1 - N \epsilon) \cdot q_\text{i} \cdot 1 + N \epsilon \cdot \lambda \cdot \frac{N-1}{N}}{(1-N \epsilon) q_\text{i} + N \epsilon \cdot \lambda} = 1 - \frac{\epsilon}{N} + \mathcal{O}(\epsilon^2),
    \label{eq:newerror}
\end{equation}
which confirms the observed $\epsilon^\prime = \frac{\epsilon}{N} + \mathcal{O}(\epsilon^2)$ scaling for Fourier transform-based distillation schemes. \\

Next, we note that this calculation directly implies that each allowed outcome with exactly 1 photon in the desired optical mode corresponds to a valid heralding pattern within the low error approximation. Eq. \ref{eq:ImplicationConjecture} shows that the probability of observing any allowed outcome $M$ either a) always decreases with increasing error, in which case the allowed outcome constitutes a distillation strategy or b) always remains zero up to $\mathcal{O}(\epsilon^2)$ in case of allowed outcomes which have $p = 0$, which is never observed in experiment in the parameter regime under consideration here. For this reason the posterior probability of there not being any errors in the state must increase after such a measurement. 

We remark that our conjecture sheds some light on why Fourier matrices repeatedly emerge as such useful tools for measuring and controlling partial distinguishability: Eq. \ref{eq:conjecture} also holds for outcomes where all photons emerge from the same mode, in any interferometer, whereas we find that for general outcomes in Haar-random interferometers, it does not hold (App. \ref{app:conjecture}). Outcomes where all photons emerge from the same mode exhibit bosonic bunching, which is a multiphoton constructive interference phenomenon that strongly modifies the output distribution of indistinguishable particles. However, in any interferometer, single-mode bunched outcomes only occur with a small probability, which moreover decreases as a function of system size \cite{Joonsuk2019bunching}. Counterintuitively, the probability of multimode bunched outcomes is not always maximized for indistinguishable photons \cite{seron2023boson}. In a Fourier matrix, in contrast, in the low error regime, our results show that all outcomes behave as if they corresponded to single-mode bunched outcomes. We leave this point for future study. 

Finally, we focus on computing the total success probability of our scheme. We carry out two approaches. First, we perform a full computation of $p := \sum q_\text{i}$, valid for all $N$. We explicitly compute $p$ by adopting the extended sample space formalism of Clifford \& Clifford \cite{clifford2018classical} to find (full derivation in App. \ref{app:heraldprob})
\begin{equation}
    p = \sum_{j=0}^{N-1} (-1)^j (j+1) \prod_{i = 1}^{j} \left(1-\frac{i}{N}\right).
    \label{eq:p}
\end{equation}

This formula holds for all $N$, but the alternating nature of the terms makes computation of an asymptote nontrivial. Numerical computations up to $N = 1000$ show that the condition $p \geq \frac{1}{4}$ holds in that regime, with an apparent asymptote at $\frac{1}{4}$. 

A second approach is concerned only with the limit of large $N$. We can make a few observations. First, we note that for any $N$, the symmetry properties of Fourier matrices mean that they obey the condition $\forall(l_i = m_i + 1 \mod N) \rightarrow p_l = p_m$, where $l$ and $m$ are mode assignment lists. In other words, if all elements of two mode assignment lists differ by one, then the probabilities are equal. Secondly, the substitution $U_{i,j} = \bar{U}_{i,j}$, where the overbar represents the complex conjugate, results in a relabeling of the modes while not altering the permanent of submatrices of $U$. These two relations define equivalence classes in the space of outputs.  Third, in the limit of large $N$, almost every allowed outcome belongs to an equivalence class that also contains allowed outcomes which correspond to photon distillation, i.e. which have exactly one photon in the mode of interest. These considerations suggest that in the limit of large $N$, imposing the restriction that there is a single photon in mode $N$ does not substantially modify the statistics of the subset of outcomes that arises as a result of imposing that condition. In other words, a sample of outcomes with this condition will have the same average value as a sample without this condition. In that case, a simple counting argument suffices to compute the limiting value of the success probability: there are $\binom{2N-1}{N}$ measurements in general of which a fraction $\frac{1}{N}$ satisfies the complementary of Eq. \ref{eq:nonforbiddenstates} \cite{tichy2010zero}. Furthermore, there are $\binom{2N-3}{N-1}$ measurements in which there is exactly 1 photon in output mode $N$, of which a fraction $\frac{1}{N}$ satisfies the complementary of Eq. \ref{eq:nonforbiddenstates} too. We find a total herald probability 
\begin{equation}
    p = \frac{\frac{1}{N}\binom{2N-3}{N-1}}{\frac{1}{N}\binom{2N-1}{N}} = \frac{1}{4-\frac{2}{N}} = \frac{1}{4} + \frac{1}{8N} + \mathcal{O}\left(\left(\frac{1}{N}\right)^2\right).
    \label{eq:p_limit_scaling}
\end{equation}
Then, it follows naturally that $\lim_{N\to\infty} p = \frac{1}{4}$. Our proof sketch explains the asymptotic behavior, but it underestimates the convergence scaling (App. \ref{app:heraldprob}, Fig. \ref{fig:p_numeric} \cite{privateCom}), demonstrating the need for future work on this topic. 

\section{Discussion}
We first focus on performance metrics in the small error limit, i.e. in the regime where Eq. \ref{eq:MultiMoylett} is valid. We are specifically interested in the number of photons required for one purified photon of with error reduction $\frac{\epsilon}{\epsilon^\prime}$. This improvement is obtained by either concatenating smaller distillation circuits or using larger distillation circuits in a single step. In either case, the photon resource cost is of the order $C = \mathcal{O}((\frac{\epsilon}{\epsilon^\prime})^\gamma)$  \cite{marshall2022distillation}. For generalized $N$-photon Fourier transform-based distillation schemes, we obtained a one-step accuracy $\frac{\epsilon}{\epsilon^\prime} = N$. Then, the scaling coefficient $\gamma$ is given by
 \begin{equation}
	\gamma := 1 + \frac{\log_2p^{-1}}{\log_2N}.
    \label{eq:gamma}
\end{equation}
 
Plugging in the asymptotic value of  $\lim_{N\to\infty} p = \frac{1}{4}$ gives: 

\begin{equation}
	\gamma \leq 1 + \frac{2}{\log_2N}
    \label{eq:ruleofthumb}
\end{equation}
 as an upper bound for the scaling coefficient.
 
\begin{figure}[h!]
    \centering
    \includegraphics[width=8.6cm, height=20cm, keepaspectratio]{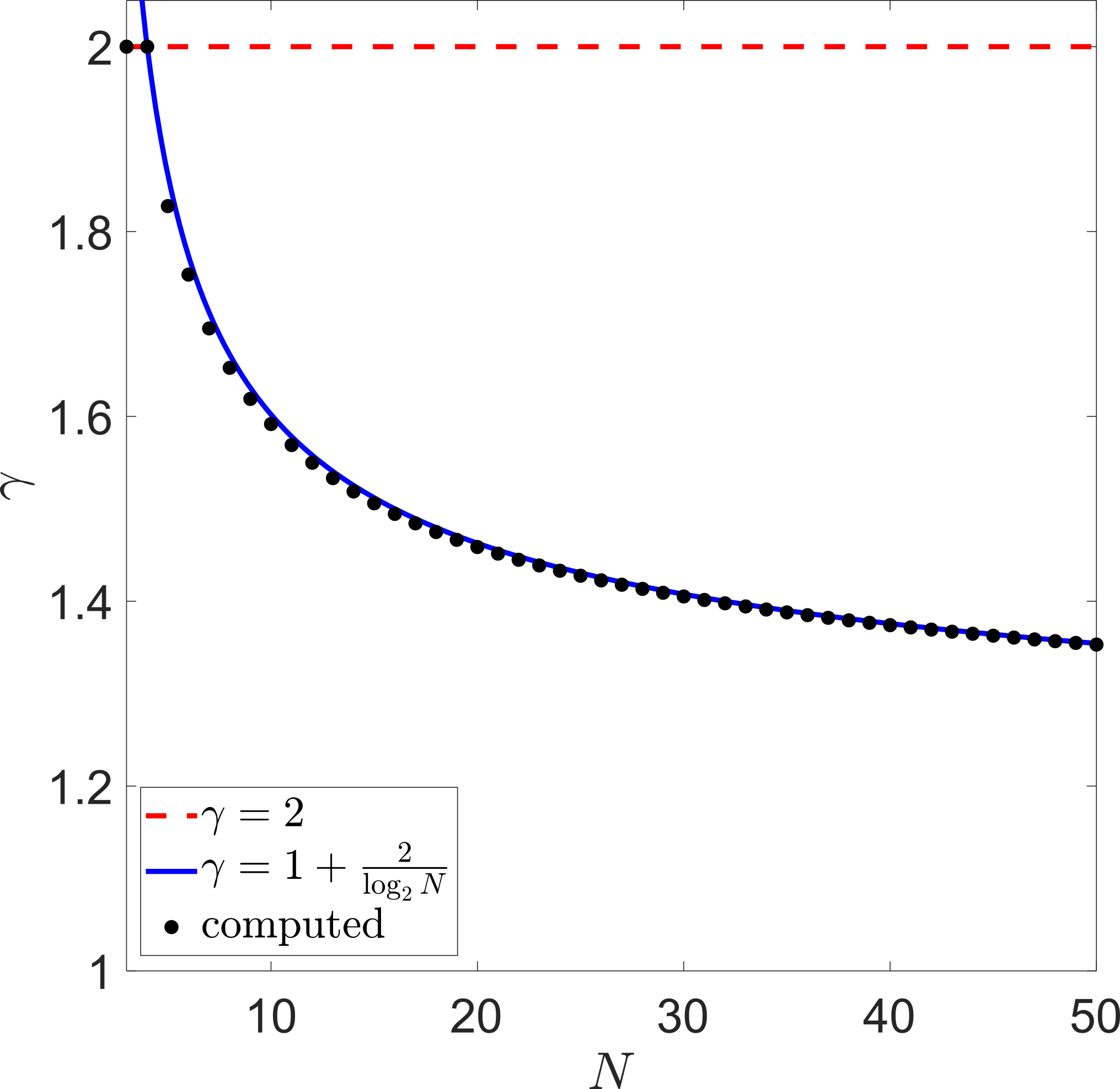}
    \caption{\textbf{Resource cost scaling coefficient for $N$-photon Fourier transform-based distillation schemes.} The figure shows the parameter $\gamma$, which characterizes the resource cost, as a function of the number of photons $N$. In the limit of small partial distinguishability error, the required number of photons to distill one purified photon is $\mathcal{O}((\frac{\epsilon}{\epsilon^\prime})^\gamma)$. The black symbols show the scaling coefficient $\gamma$ as computed via direct computation. The blue line shows an upper bound (Eq. \ref{eq:ruleofthumb}, blue line) which only slightly overestimates the exact calculations. $N = 3,4$ follow the quadratic scaling law found in Ref. \cite{marshall2022distillation} (red dashed line).}
    \label{fig:gamma}
\end{figure}

 Fig. \ref{fig:gamma} shows $\gamma$ as a function of $N$, both using an exact computation of $p$ from Eq. \ref{eq:p}, and using the upper bound. Eq. \ref{eq:ruleofthumb} only slightly overestimates the required number of photons for $N \geq 5$. Eq. \ref{eq:ruleofthumb} implies a transition from quadratic to linear scaling for sufficiently large $N$-photon Fourier transform-based distillation schemes. This transition ultimately arises because our results show that the optimal strategy in the small-error regime is to take as many photons as are necessary to achieve the error reduction target and interfere them in a single distillation step. Because that distillation step succeeds with constant probability independent of the system size, this is the optimal strategy. In particular, we find
 \begin{equation}
     \left(\frac{\epsilon}{\epsilon^\prime}\right)^\gamma \leq \left(\frac{\epsilon}{\epsilon^\prime}\right)^{\log_N{4}}\cdot \left(\frac{\epsilon}{\epsilon^\prime}\right).
 \end{equation}
Since $\epsilon / \epsilon^\prime = N + \mathcal{O}(\epsilon^2)$, we find a linear resource cost scaling
\begin{equation}
    C \leq \mathcal{O}\left(4\left(\frac{\epsilon}{\epsilon^\prime}\right) + \mathcal{O}\left(\epsilon^3\right) \right). 
\end{equation}

Next, we discuss the applicability of our results in the regime where Eq. \ref{eq:MultiMoylett} does not hold. In this regime, the error reduction depends on the specific heralding pattern, and therefore each heralding pattern has a different error threshold and error reduction, as as can be seen in Fig. \ref{fig:fig3_resulterror}. It is therefore necessary to know both the input error of the photon and the function that describes the error reduction as a function of input error and herald pattern. Since these functions are polynomials in the error, but computing the coefficients of these polynomials is a computationally expensive task, and since very similar computations would have to occur many times in the same clock cycle in a photonic quantum computer, it makes sense to pre-compute the coefficients of these polynomials, or even to create a lookup table \footnote{Pre-computed data for $N = 3$ to $N = 10$ can be found in \cite{dataSet}}. Furthermore, in this regime, it may be optimal to perform multiple distillation steps, since the convergence to a constant success probability cannot be guaranteed. Since in this regime, the error reduction cannot be predicted before each distillation step, the distillation protocol will have to be adaptive, relying on multiple possible routing schemes, which may themselves be pre-computed, to route the photons.

\section{Implementation in photonic quantum computers}

We conclude our work by discussing how our findings will find application in large-scale photonic quantum computers. To achieve large-scale quantum computation, errors in the computation must be reduced to a level that on average, much less than one logical error occurs during a computation. Conventionally, this is done by a combination of low intrinsic error rates on the individual components and quantum error correction. The threshold theorem states that for each error correcting code, there exists a threshold $p_{\text{th}}$ such that if $p_{\text{error}} < p_{\text{th}}$, the logical error $p_{\text{L}}$ can be arbitrarily reduced \cite{aharonov1997fault}. For the purposes of our discussion, low error rates $p_{\text{error}}$ correspond to better quality photon sources. We address the question of how photon distillation fits in this picture. 

First, we discuss the relation between photon distillation and improvements in photon source quality. Photonic quantum computing will require mass manufacturing of photon sources in scalable foundry processes. Therefore, improvement in photon indistinguishability through engineering is ultimately limited by uncontrollable fabrication process errors \cite{alexander2024manufacturable}. Overcoming these limitations by post-fabrication calibration is limited and costly, because accurately identifying the exact cause of partial distinguishability is essential. Both photon distillation and quantum error correction have the advantage that it is not required to identify the exact cause of partial distinguishability. Instead, for photon distillation, the use of quantum interference of single photon wavefunctions provides simultaneous filtering in all internal error modes. 

The more direct comparison is therefore with quantum error correction (QEC), where once the engineering of photon sources has reached a natural end point, the choice is to deploy QEC, photon distillation, or a combination of both. There are two regimes of interest: the regime where the probability of error is lower than the error threshold of the error correcting code, and the regime where it is above the threshold. We will work under the assumption that the distinguishability error $\epsilon$ is such that we are operating in the small-error regime of photon distillation. We will see at the end of this section that this assumption is justified for most practically relevant cases, and discuss how our results are modified if it is not. 

We will first consider the regime below threshold. Here, the key consideration is resource count: given photon sources of a given quality, and a given target logical error rate, what combination of QEC and photon distillation results in the fewest photon sources required to create a single logical qubit? From the perspective of QEC, partial distinguishability errors are known to cause Pauli stabilizer errors, to reduce the purity of resource states, and to force encoded quantum information out of the code space \cite{rohde2006error, sparrow2017quantum, saied2024advancing}. In our exploratory analysis of the asymptotically low $\epsilon$ regime, these errors are accounted for in $p_{\text{error}}$. For surface codes, the logical error $p_{\text{L}}$ is reduced by increasing the code distance $d$ \cite{fowler2012surface}: 
\begin{equation}
	p_{\text{L}} = a \cdot \left( \frac{p_{\text{error}}}{ p_{\text{th}}} \right) ^\frac{d}{2},
	\label{eq:QECsurfacecode}
\end{equation}
where $a$ is a proportionality constant and $d$ is assumed to be even. The code distance $d$ is a measure for the $d^2$ resource states needed to form a surface code patch to encode one logical qubit. Considering a fixed photon number overhead $b$ required to create a resource state, the estimated  resource cost for a logical qubit is $C_{\text{L}} = b \cdot d^2$ photons. 

\begin{figure}[h!]
    \centering
    \includegraphics[width=8.6cm, height=20cm, keepaspectratio]{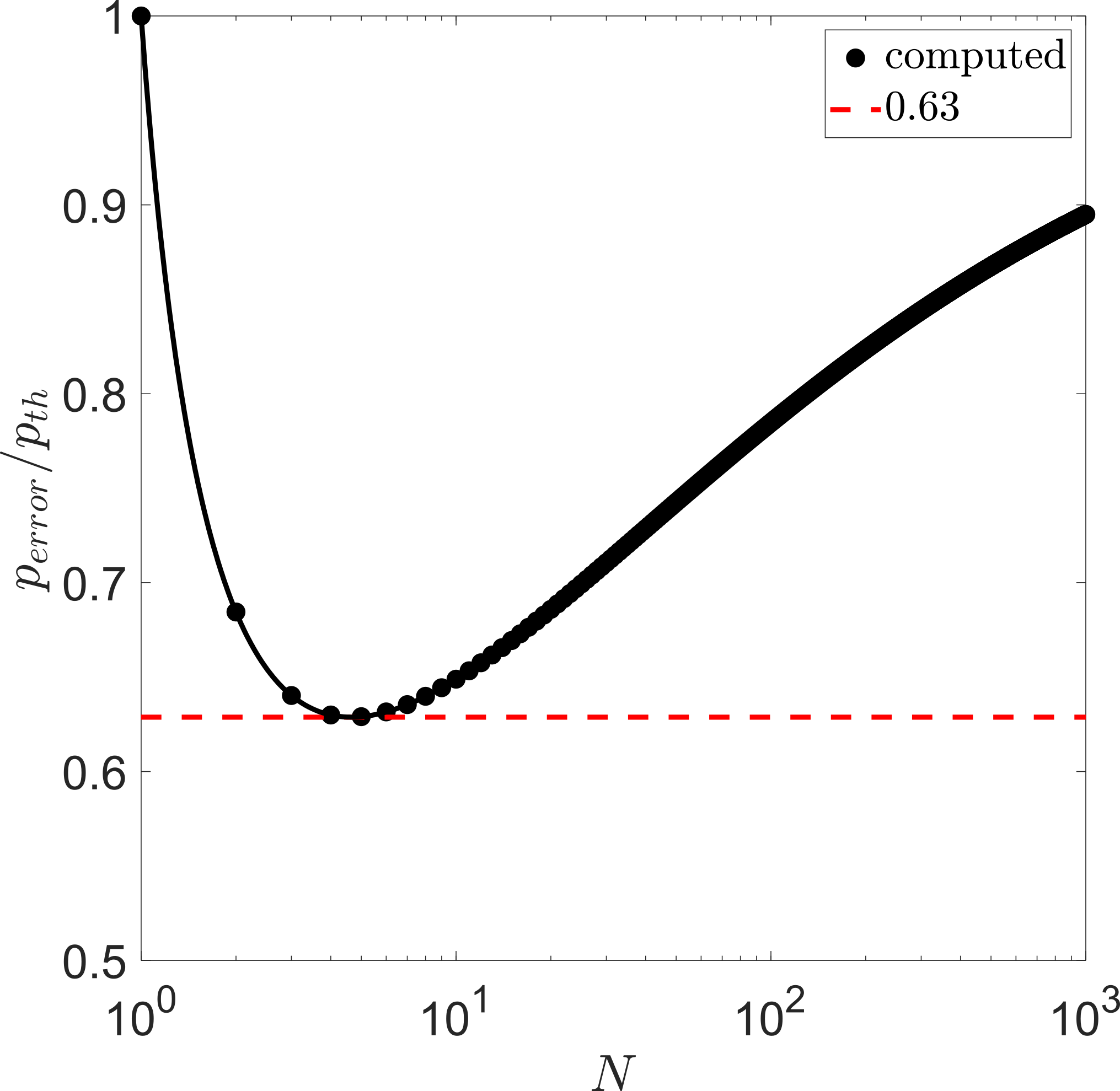}
    \caption{\textbf{Photon distillation resource allocation trade-off.} This figure shows the trade-off between allocating photon resources for either error mitigation (by means of photon distillation) or enhancing the quantum error-correction code distance.  The use of $N$-photon distillation schemes is advantageous when the initial error rate fraction $p_{\text{error}}/p_{\text{th}}$ exceeds the black curve. Generally, the allocation of photon resources to photon distillation is effective when $p_{\text{error}}/p_{\text{th}} > 0.63$.}
    \label{fig:QEC_tradeoff}
\end{figure}

Linear optical quantum gates can be modeled as the interference of a mixed state of (in)distinguishable photons, followed by measurement and optical switching based on the outcome of that measurement. We make the conservative approximation that these gates fail if even a single one of the injected photons is fully distinguishable. Inspired by Eq. \ref{eq:MultiMoylett}, we therefore propose to model $p_{\text{error}} = \mathcal{O}(\epsilon)$ \footnote{This is a conservative model: if $p_{\text{error}} = \mathcal{O}(\epsilon^k)$ for integer $k \geq 2$, the impact of photon distillation will be even greater, as the factor $1/N$ in Eq. \ref{eq:inequalityQEC} will be replaced by $1/N^k$.}. Photon distillation in the limit of small errors reduces the physical error to $p_{\text{error}} / N$ at the cost of an additional fixed photon number overhead factor of $4N$. Simultaneously, the required code distance to achieve the same logical error as in the scenario without photon distillation is reduced. Specifically, we find that while using photon distillation, the photon resource cost for one logical qubit with similar logical error is
\begin{equation}
C_{\text{L}}^{\text{dist.}} = C_{\text{L}} \cdot \left(  \frac{2\sqrt{N} \log(\frac{p_{\text{error}}}{p_{\text{th}}} ) }{  \log(\frac{p_{\text{error}}}{p_{\text{th}}} ) - \log(N)}          \right)^2.
\label{eq:C_cost_distillation}
\end{equation}
Noticeably, we find that photon distillation outperforms surface code QEC ($C_{\text{L}}^{\text{dist.}}< C_{\text{L}}$) if
\begin{equation}
	\frac{p_{\text{error}}}{p_{\text{th}}} > \left( \frac{1}{N} \right) ^{\frac{1}{2\sqrt{N} – 1}}.
    \label{eq:inequalityQEC}
\end{equation}
In Fig. \ref{fig:QEC_tradeoff}, we observe that photon distillation is viable for $p_{\text{error}}/p_{\text{th}} > 0.63$, showing that a regime exists where a combination of photon distillation and QEC has reduced resource costs compared to a strategy where only QEC is employed.  

Next, we consider the regime above the QEC threshold. Making a direct comparison between photon distillation and quantum error correction requires establishing a correspondence between logical errors $p_{\textrm{error}}$ and photon errors $\epsilon$. We note that since a photonic quantum computer consumes a fixed number of photons per qubit per computational step, these are both per-qubit, per-step quantities. However, a precise correspondence will depend on the specific architecture used and might even vary depending on where in the computer the error occurs. We note, however, that all photonic quantum gates operate on a small number of $N_{\textrm{gate}} = \mathcal{O}(1)$ photons. Following the approximation that a single error photon induces gate failure, this means that $p_{\textrm{error}}/\epsilon = N_{\textrm{gate}} \approx 1$. This result is consistent with studies of distinguishability errors in specific cases e.g., in dephasing-only error model for cluster states in Ref. \cite{rohde2006error}. Here, $p_{\text{error}} = \frac{1}{2} - \frac{(1-\epsilon)^2}{1 + (1-\epsilon)^2} = \frac{1}{2}\epsilon + \mathcal{O}(\epsilon^2)$.

Establishing this correspondence allows us to make a direct comparison: the error threshold for photon distillation scales as $\epsilon^* \approx 1/N^{0.61}$ \cite{saied2024general}, giving results of $\epsilon \simeq 0.2-0.35$, for the regime where Fig. 5 suggests photon distillation is most resource-efficient. In contrast, thresholds for QEC are $p_{\text{th}} = \mathcal{O}(10^{-4} – 10^{-1})$ \cite{bartolucci2023fusion,rohde2006error}, with FBQC having a threshold of up to 1.07\% Pauli error \cite{bartolucci2023fusion}. 

Our results strongly suggest that even for a very pessimistic estimate of $N_{\textrm{gate}}$ of, say, $N_{\textrm{gate}} = 10$, there exists a regime where the combination of photon distillation followed by QEC has a higher threshold than only QEC. Furthermore, we note that many existing photon sources operate already below the threshold for photon distillation \cite{saied2024general}. 

Finally, we return to the question of whether the use of the small-error limit is justified in these calculations. For our result below threshold, it is clear that this is the case if we, for example, assume the FBQC threshold of 1.07\% error. For the results above the QEC threshold, as noted above, even with a pessimistic assumption about $N_{\textrm{gate}}$, we can show that the theshold of the combined scheme exceeds that of bare QEC. However, a precise calculation of the threshold of a combined photon distillation plus QEC scheme is beyond the present work, as this requires both a better understanding of the high-error regime as well as a better understanding of the relationship between partial distinguishability errors and logical errors in photonic quantum computation. Furthermore, each additional component added to photonic quantum computing architectures introduces new losses, and photon distillation gates are no exception. Therefore, photon distillation gates inherently reduce indistinguishability errors, but at the expense of increased loss errors. In App. \ref{app:freelunch}, we discuss the error scaling of realistic photon distillation gates in the presence of loss. We defer the full generalization of including other error types for future work. 

\section{Conclusion}
In conclusion, we have used a combination of numerical and theoretical tools to shown that large Fourier transform-based schemes can be used to distill indistinguishable photons from imperfect sources. For this purpose, we explicitly simulated the error reduction for the $N = 5$ photon Fourier transform where we showed that there are different paths towards perfect photon indistinguishability. Within the limit of low partial distinguishability error, we have shown modulo a conjecture that the error reduction scales linearly with the number of photons employed in the distillation scheme $N$, while we have given numerical and theoretical evidence that the success probability tends to a constant, giving rise to linear scaling in the number of photons required to achieve a given error reduction.

Nevertheless, applicability of our work strongly depends on the availability of high-fidelity programmable photonic processors and fast optical switches. However, we note that processor size and fidelity of linear optical circuits are constantly improving \cite{carolan2015universal,harris2017quantum,sparrow2018simulating,wang2020integrated,taballione2021universal, taballione202320}. Our scalable scheme contributes to the development of adaptive photon distillation circuits, where concatenation strategies are dependent on the actual herald measurements of each distillation step. While current advances in single photon source engineering continue to improve photon indistinguishability \cite{tomm2021bright,ollivier2021hong,becher20232023,ding2023high}, we anticipate that photon distillation will aid in the task of creating the millions of almost fully indistinguishable photons necessary for photonic quantum computing. We have argued that a higher error threshold can be achieved by complementing conventional QEC methods with photon distillation, and that a regime exists below threshold where a combination of photon distillation and QEC outperforms bare QEC in terms of resource costs at constant performance. These results are strong evidence that our protocol will find application in large-scale photonic quantum computers. 

\section*{Acknowledgements}
We thank J. Timmerhuis for scientific discussions. We thank J. Saied and J. Marshall for helpful comments on our first draft of the manuscript. This research is supported by the Photonic Integrated Technology Center (PITC). This project has received funding from PhotonDelta National Growth Fund programme. This publication is part of the project At the Quantum Edge (VI.Vidi.223.075) of the research programme VIDI which is financed by the Dutch Research Council (NWO).

\section*{Competing interests}
F.H.B.S, B.K.S, and J.J.R. are the inventors on a relevant patent application submitted by the University of Twente (Dutch patent application no. 2037489). The authors declare no other competing interests.

\bibliography{refs.bib}

\begin{thebibliography}{64}%
\makeatletter
\providecommand \@ifxundefined [1]{%
 \@ifx{#1\undefined}
}%
\providecommand \@ifnum [1]{%
 \ifnum #1\expandafter \@firstoftwo
 \else \expandafter \@secondoftwo
 \fi
}%
\providecommand \@ifx [1]{%
 \ifx #1\expandafter \@firstoftwo
 \else \expandafter \@secondoftwo
 \fi
}%
\providecommand \natexlab [1]{#1}%
\providecommand \enquote  [1]{``#1''}%
\providecommand \bibnamefont  [1]{#1}%
\providecommand \bibfnamefont [1]{#1}%
\providecommand \citenamefont [1]{#1}%
\providecommand \href@noop [0]{\@secondoftwo}%
\providecommand \href [0]{\begingroup \@sanitize@url \@href}%
\providecommand \@href[1]{\@@startlink{#1}\@@href}%
\providecommand \@@href[1]{\endgroup#1\@@endlink}%
\providecommand \@sanitize@url [0]{\catcode `\\12\catcode `\$12\catcode `\&12\catcode `\#12\catcode `\^12\catcode `\_12\catcode `\%12\relax}%
\providecommand \@@startlink[1]{}%
\providecommand \@@endlink[0]{}%
\providecommand \url  [0]{\begingroup\@sanitize@url \@url }%
\providecommand \@url [1]{\endgroup\@href {#1}{\urlprefix }}%
\providecommand \urlprefix  [0]{URL }%
\providecommand \Eprint [0]{\href }%
\providecommand \doibase [0]{https://doi.org/}%
\providecommand \selectlanguage [0]{\@gobble}%
\providecommand \bibinfo  [0]{\@secondoftwo}%
\providecommand \bibfield  [0]{\@secondoftwo}%
\providecommand \translation [1]{[#1]}%
\providecommand \BibitemOpen [0]{}%
\providecommand \bibitemStop [0]{}%
\providecommand \bibitemNoStop [0]{.\EOS\space}%
\providecommand \EOS [0]{\spacefactor3000\relax}%
\providecommand \BibitemShut  [1]{\csname bibitem#1\endcsname}%
\let\auto@bib@innerbib\@empty
\bibitem [{\citenamefont {Knill}\ \emph {et~al.}(2001)\citenamefont {Knill}, \citenamefont {Laflamme},\ and\ \citenamefont {Milburn}}]{knill2001scheme}%
  \BibitemOpen
  \bibfield  {author} {\bibinfo {author} {\bibfnamefont {E.}~\bibnamefont {Knill}}, \bibinfo {author} {\bibfnamefont {R.}~\bibnamefont {Laflamme}},\ and\ \bibinfo {author} {\bibfnamefont {G.~J.}\ \bibnamefont {Milburn}},\ }\bibfield  {title} {\bibinfo {title} {A scheme for efficient quantum computation with linear optics},\ }\href {https://doi.org/https://doi.org/10.1038/35051009} {\bibfield  {journal} {\bibinfo  {journal} {Nature}\ }\textbf {\bibinfo {volume} {409}},\ \bibinfo {pages} {46} (\bibinfo {year} {2001})}\BibitemShut {NoStop}%
\bibitem [{\citenamefont {Browne}\ and\ \citenamefont {Rudolph}(2005)}]{browne2005resource}%
  \BibitemOpen
  \bibfield  {author} {\bibinfo {author} {\bibfnamefont {D.~E.}\ \bibnamefont {Browne}}\ and\ \bibinfo {author} {\bibfnamefont {T.}~\bibnamefont {Rudolph}},\ }\bibfield  {title} {\bibinfo {title} {Resource-efficient linear optical quantum computation},\ }\href {https://doi.org/https://doi.org/10.1103/PhysRevLett.95.010501} {\bibfield  {journal} {\bibinfo  {journal} {Physical Review Letters}\ }\textbf {\bibinfo {volume} {95}},\ \bibinfo {pages} {010501} (\bibinfo {year} {2005})}\BibitemShut {NoStop}%
\bibitem [{\citenamefont {Rudolph}(2017)}]{rudolph2017optimistic}%
  \BibitemOpen
  \bibfield  {author} {\bibinfo {author} {\bibfnamefont {T.}~\bibnamefont {Rudolph}},\ }\bibfield  {title} {\bibinfo {title} {Why i am optimistic about the silicon-photonic route to quantum computing},\ }\href {https://doi.org/https://doi.org/10.1063/1.4976737} {\bibfield  {journal} {\bibinfo  {journal} {APL Photonics}\ }\textbf {\bibinfo {volume} {2}},\ \bibinfo {pages} {030901} (\bibinfo {year} {2017})}\BibitemShut {NoStop}%
\bibitem [{\citenamefont {Bartolucci}\ \emph {et~al.}(2023)\citenamefont {Bartolucci}, \citenamefont {Birchall}, \citenamefont {Bombin}, \citenamefont {Cable}, \citenamefont {Dawson}, \citenamefont {Gimeno-Segovia}, \citenamefont {Johnston}, \citenamefont {Kieling}, \citenamefont {Nickerson}, \citenamefont {Pant} \emph {et~al.}}]{bartolucci2023fusion}%
  \BibitemOpen
  \bibfield  {author} {\bibinfo {author} {\bibfnamefont {S.}~\bibnamefont {Bartolucci}}, \bibinfo {author} {\bibfnamefont {P.}~\bibnamefont {Birchall}}, \bibinfo {author} {\bibfnamefont {H.}~\bibnamefont {Bombin}}, \bibinfo {author} {\bibfnamefont {H.}~\bibnamefont {Cable}}, \bibinfo {author} {\bibfnamefont {C.}~\bibnamefont {Dawson}}, \bibinfo {author} {\bibfnamefont {M.}~\bibnamefont {Gimeno-Segovia}}, \bibinfo {author} {\bibfnamefont {E.}~\bibnamefont {Johnston}}, \bibinfo {author} {\bibfnamefont {K.}~\bibnamefont {Kieling}}, \bibinfo {author} {\bibfnamefont {N.}~\bibnamefont {Nickerson}}, \bibinfo {author} {\bibfnamefont {M.}~\bibnamefont {Pant}}, \emph {et~al.},\ }\bibfield  {title} {\bibinfo {title} {Fusion-based quantum computation},\ }\href {https://doi.org/https://doi.org/10.1038/s41467-023-36493-1} {\bibfield  {journal} {\bibinfo  {journal} {Nature Communications}\ }\textbf {\bibinfo {volume} {14}},\ \bibinfo {pages} {912} (\bibinfo {year} {2023})}\BibitemShut {NoStop}%
\bibitem [{\citenamefont {Chen}\ \emph {et~al.}(2024)\citenamefont {Chen}, \citenamefont {Peng}, \citenamefont {Guo}, \citenamefont {Gu}, \citenamefont {Ding}, \citenamefont {Liu}, \citenamefont {You}, \citenamefont {Qin}, \citenamefont {Wang}, \citenamefont {He} \emph {et~al.}}]{chen2023heralded}%
  \BibitemOpen
  \bibfield  {author} {\bibinfo {author} {\bibfnamefont {S.}~\bibnamefont {Chen}}, \bibinfo {author} {\bibfnamefont {L.-C.}\ \bibnamefont {Peng}}, \bibinfo {author} {\bibfnamefont {Y.-P.}\ \bibnamefont {Guo}}, \bibinfo {author} {\bibfnamefont {X.-M.}\ \bibnamefont {Gu}}, \bibinfo {author} {\bibfnamefont {X.}~\bibnamefont {Ding}}, \bibinfo {author} {\bibfnamefont {R.-Z.}\ \bibnamefont {Liu}}, \bibinfo {author} {\bibfnamefont {X.}~\bibnamefont {You}}, \bibinfo {author} {\bibfnamefont {J.}~\bibnamefont {Qin}}, \bibinfo {author} {\bibfnamefont {Y.-F.}\ \bibnamefont {Wang}}, \bibinfo {author} {\bibfnamefont {Y.-M.}\ \bibnamefont {He}}, \emph {et~al.},\ }\bibfield  {title} {\bibinfo {title} {Heralded three-photon entanglement from a single-photon source on a photonic chip},\ }\href {https://doi.org/10.1103/PhysRevLett.132.130603} {\bibfield  {journal} {\bibinfo  {journal} {Physical Review Letters}\ }\textbf {\bibinfo {volume} {132}},\ \bibinfo {pages} {130603} (\bibinfo {year} {2024})}\BibitemShut {NoStop}%
\bibitem [{\citenamefont {Cao}\ \emph {et~al.}(2024)\citenamefont {Cao}, \citenamefont {Hansen}, \citenamefont {Giorgino}, \citenamefont {Carosini}, \citenamefont {Zah{\'a}lka}, \citenamefont {Zilk}, \citenamefont {Loredo},\ and\ \citenamefont {Walther}}]{cao2023photonic}%
  \BibitemOpen
  \bibfield  {author} {\bibinfo {author} {\bibfnamefont {H.}~\bibnamefont {Cao}}, \bibinfo {author} {\bibfnamefont {L.}~\bibnamefont {Hansen}}, \bibinfo {author} {\bibfnamefont {F.}~\bibnamefont {Giorgino}}, \bibinfo {author} {\bibfnamefont {L.}~\bibnamefont {Carosini}}, \bibinfo {author} {\bibfnamefont {P.}~\bibnamefont {Zah{\'a}lka}}, \bibinfo {author} {\bibfnamefont {F.}~\bibnamefont {Zilk}}, \bibinfo {author} {\bibfnamefont {J.}~\bibnamefont {Loredo}},\ and\ \bibinfo {author} {\bibfnamefont {P.}~\bibnamefont {Walther}},\ }\bibfield  {title} {\bibinfo {title} {Photonic source of heralded greenberger-horne-zeilinger states},\ }\href {https://doi.org/https://doi.org/10.1103/PhysRevLett.132.130604} {\bibfield  {journal} {\bibinfo  {journal} {Physical Review Letters}\ }\textbf {\bibinfo {volume} {132}},\ \bibinfo {pages} {130604} (\bibinfo {year} {2024})}\BibitemShut {NoStop}%
\bibitem [{\citenamefont {Maring}\ \emph {et~al.}(2024)\citenamefont {Maring}, \citenamefont {Fyrillas}, \citenamefont {Pont}, \citenamefont {Ivanov}, \citenamefont {Stepanov}, \citenamefont {Margaria}, \citenamefont {Hease}, \citenamefont {Pishchagin}, \citenamefont {Lema{\^\i}tre}, \citenamefont {Sagnes} \emph {et~al.}}]{maring2023general}%
  \BibitemOpen
  \bibfield  {author} {\bibinfo {author} {\bibfnamefont {N.}~\bibnamefont {Maring}}, \bibinfo {author} {\bibfnamefont {A.}~\bibnamefont {Fyrillas}}, \bibinfo {author} {\bibfnamefont {M.}~\bibnamefont {Pont}}, \bibinfo {author} {\bibfnamefont {E.}~\bibnamefont {Ivanov}}, \bibinfo {author} {\bibfnamefont {P.}~\bibnamefont {Stepanov}}, \bibinfo {author} {\bibfnamefont {N.}~\bibnamefont {Margaria}}, \bibinfo {author} {\bibfnamefont {W.}~\bibnamefont {Hease}}, \bibinfo {author} {\bibfnamefont {A.}~\bibnamefont {Pishchagin}}, \bibinfo {author} {\bibfnamefont {A.}~\bibnamefont {Lema{\^\i}tre}}, \bibinfo {author} {\bibfnamefont {I.}~\bibnamefont {Sagnes}}, \emph {et~al.},\ }\bibfield  {title} {\bibinfo {title} {A versatile single-photon-based quantum computing platform},\ }\href {https://doi.org/https://doi.org/10.1038/s41566-024-01403-4} {\bibfield  {journal} {\bibinfo  {journal} {Nature Photonics}\ ,\ \bibinfo {pages} {1}} (\bibinfo {year} {2024})}\BibitemShut {NoStop}%
\bibitem [{\citenamefont {Mandel}(1991)}]{mandel1991coherence}%
  \BibitemOpen
  \bibfield  {author} {\bibinfo {author} {\bibfnamefont {L.}~\bibnamefont {Mandel}},\ }\bibfield  {title} {\bibinfo {title} {Coherence and indistinguishability},\ }\href {https://doi.org/https://doi.org/10.1364/OL.16.001882} {\bibfield  {journal} {\bibinfo  {journal} {Optics Letters}\ }\textbf {\bibinfo {volume} {16}},\ \bibinfo {pages} {1882} (\bibinfo {year} {1991})}\BibitemShut {NoStop}%
\bibitem [{\citenamefont {Shchesnovich}(2014)}]{shchesnovich2014sufficient}%
  \BibitemOpen
  \bibfield  {author} {\bibinfo {author} {\bibfnamefont {V.}~\bibnamefont {Shchesnovich}},\ }\bibfield  {title} {\bibinfo {title} {Sufficient condition for the mode mismatch of single photons for scalability of the boson-sampling computer},\ }\href {https://doi.org/https://doi.org/10.1103/PhysRevA.89.022333} {\bibfield  {journal} {\bibinfo  {journal} {Physical Review A}\ }\textbf {\bibinfo {volume} {89}},\ \bibinfo {pages} {022333} (\bibinfo {year} {2014})}\BibitemShut {NoStop}%
\bibitem [{\citenamefont {Lal}\ \emph {et~al.}(2022)\citenamefont {Lal}, \citenamefont {Mishra},\ and\ \citenamefont {Singh}}]{lal2022indistinguishable}%
  \BibitemOpen
  \bibfield  {author} {\bibinfo {author} {\bibfnamefont {N.}~\bibnamefont {Lal}}, \bibinfo {author} {\bibfnamefont {S.}~\bibnamefont {Mishra}},\ and\ \bibinfo {author} {\bibfnamefont {R.}~\bibnamefont {Singh}},\ }\bibfield  {title} {\bibinfo {title} {Indistinguishable photons},\ }\bibfield  {journal} {\bibinfo  {journal} {AVS Quantum Science}\ }\textbf {\bibinfo {volume} {4}},\ \href {https://doi.org/https://doi.org/10.1116/5.0083968} {https://doi.org/10.1116/5.0083968} (\bibinfo {year} {2022})\BibitemShut {NoStop}%
\bibitem [{\citenamefont {Flamini}\ \emph {et~al.}(2018)\citenamefont {Flamini}, \citenamefont {Spagnolo},\ and\ \citenamefont {Sciarrino}}]{flamini2018photonic}%
  \BibitemOpen
  \bibfield  {author} {\bibinfo {author} {\bibfnamefont {F.}~\bibnamefont {Flamini}}, \bibinfo {author} {\bibfnamefont {N.}~\bibnamefont {Spagnolo}},\ and\ \bibinfo {author} {\bibfnamefont {F.}~\bibnamefont {Sciarrino}},\ }\bibfield  {title} {\bibinfo {title} {Photonic quantum information processing: a review},\ }\href@noop {} {\bibfield  {journal} {\bibinfo  {journal} {Reports on Progress in Physics}\ }\textbf {\bibinfo {volume} {82}},\ \bibinfo {pages} {016001} (\bibinfo {year} {2018})}\BibitemShut {NoStop}%
\bibitem [{\citenamefont {Hong}\ \emph {et~al.}(1987)\citenamefont {Hong}, \citenamefont {Ou},\ and\ \citenamefont {Mandel}}]{hong1987measurement}%
  \BibitemOpen
  \bibfield  {author} {\bibinfo {author} {\bibfnamefont {C.-K.}\ \bibnamefont {Hong}}, \bibinfo {author} {\bibfnamefont {Z.-Y.}\ \bibnamefont {Ou}},\ and\ \bibinfo {author} {\bibfnamefont {L.}~\bibnamefont {Mandel}},\ }\bibfield  {title} {\bibinfo {title} {Measurement of subpicosecond time intervals between two photons by interference},\ }\href {https://doi.org/https://doi.org/10.1103/PhysRevLett.59.2044} {\bibfield  {journal} {\bibinfo  {journal} {Physical Review Letters}\ }\textbf {\bibinfo {volume} {59}},\ \bibinfo {pages} {2044} (\bibinfo {year} {1987})}\BibitemShut {NoStop}%
\bibitem [{\citenamefont {Kok}\ \emph {et~al.}(2007)\citenamefont {Kok}, \citenamefont {Munro}, \citenamefont {Nemoto}, \citenamefont {Ralph}, \citenamefont {Dowling},\ and\ \citenamefont {Milburn}}]{kok2007linear}%
  \BibitemOpen
  \bibfield  {author} {\bibinfo {author} {\bibfnamefont {P.}~\bibnamefont {Kok}}, \bibinfo {author} {\bibfnamefont {W.~J.}\ \bibnamefont {Munro}}, \bibinfo {author} {\bibfnamefont {K.}~\bibnamefont {Nemoto}}, \bibinfo {author} {\bibfnamefont {T.~C.}\ \bibnamefont {Ralph}}, \bibinfo {author} {\bibfnamefont {J.~P.}\ \bibnamefont {Dowling}},\ and\ \bibinfo {author} {\bibfnamefont {G.~J.}\ \bibnamefont {Milburn}},\ }\bibfield  {title} {\bibinfo {title} {Linear optical quantum computing with photonic qubits},\ }\href {https://doi.org/https://doi.org/10.1103/RevModPhys.79.135} {\bibfield  {journal} {\bibinfo  {journal} {Reviews of modern physics}\ }\textbf {\bibinfo {volume} {79}},\ \bibinfo {pages} {135} (\bibinfo {year} {2007})}\BibitemShut {NoStop}%
\bibitem [{\citenamefont {O'brien}(2007)}]{o2007optical}%
  \BibitemOpen
  \bibfield  {author} {\bibinfo {author} {\bibfnamefont {J.~L.}\ \bibnamefont {O'brien}},\ }\bibfield  {title} {\bibinfo {title} {Optical quantum computing},\ }\href {https://doi.org/https://doi.org/10.1126/science.1142892} {\bibfield  {journal} {\bibinfo  {journal} {Science}\ }\textbf {\bibinfo {volume} {318}},\ \bibinfo {pages} {1567} (\bibinfo {year} {2007})}\BibitemShut {NoStop}%
\bibitem [{\citenamefont {Varnava}\ \emph {et~al.}(2008)\citenamefont {Varnava}, \citenamefont {Browne},\ and\ \citenamefont {Rudolph}}]{varnava2008good}%
  \BibitemOpen
  \bibfield  {author} {\bibinfo {author} {\bibfnamefont {M.}~\bibnamefont {Varnava}}, \bibinfo {author} {\bibfnamefont {D.~E.}\ \bibnamefont {Browne}},\ and\ \bibinfo {author} {\bibfnamefont {T.}~\bibnamefont {Rudolph}},\ }\bibfield  {title} {\bibinfo {title} {How good must single photon sources and detectors be for efficient linear optical quantum computation?},\ }\href {https://doi.org/https://doi.org/10.1103/PhysRevLett.100.060502} {\bibfield  {journal} {\bibinfo  {journal} {Physical Review Letters}\ }\textbf {\bibinfo {volume} {100}},\ \bibinfo {pages} {060502} (\bibinfo {year} {2008})}\BibitemShut {NoStop}%
\bibitem [{\citenamefont {Slussarenko}\ and\ \citenamefont {Pryde}(2019)}]{slussarenko2019photonic}%
  \BibitemOpen
  \bibfield  {author} {\bibinfo {author} {\bibfnamefont {S.}~\bibnamefont {Slussarenko}}\ and\ \bibinfo {author} {\bibfnamefont {G.~J.}\ \bibnamefont {Pryde}},\ }\bibfield  {title} {\bibinfo {title} {Photonic quantum information processing: A concise review},\ }\bibfield  {journal} {\bibinfo  {journal} {Applied Physics Reviews}\ }\textbf {\bibinfo {volume} {6}},\ \href {https://doi.org/https://doi.org/10.1063/1.5115814} {https://doi.org/10.1063/1.5115814} (\bibinfo {year} {2019})\BibitemShut {NoStop}%
\bibitem [{\citenamefont {Sparrow}(2017)}]{sparrow2017quantum}%
  \BibitemOpen
  \bibfield  {author} {\bibinfo {author} {\bibfnamefont {C.}~\bibnamefont {Sparrow}},\ }\emph {\bibinfo {title} {Quantum interference in universal linear optical devices for quantum computation and simulation}},\ \href {https://doi.org/https://doi.org/10.25560/67638} {Ph.D. thesis} (\bibinfo {year} {2017})\BibitemShut {NoStop}%
\bibitem [{\citenamefont {Marshall}(2022)}]{marshall2022distillation}%
  \BibitemOpen
  \bibfield  {author} {\bibinfo {author} {\bibfnamefont {J.}~\bibnamefont {Marshall}},\ }\bibfield  {title} {\bibinfo {title} {Distillation of indistinguishable photons},\ }\href {https://doi.org/https://doi.org/10.1103/PhysRevLett.129.213601} {\bibfield  {journal} {\bibinfo  {journal} {Physical Review Letters}\ }\textbf {\bibinfo {volume} {129}},\ \bibinfo {pages} {213601} (\bibinfo {year} {2022})}\BibitemShut {NoStop}%
\bibitem [{\citenamefont {Faurby}\ \emph {et~al.}(2024)\citenamefont {Faurby}, \citenamefont {Carosini}, \citenamefont {Cao}, \citenamefont {Sund}, \citenamefont {Hansen}, \citenamefont {Giorgino}, \citenamefont {Villadsen}, \citenamefont {Van Den~Hoven}, \citenamefont {Lodahl}, \citenamefont {Paesani} \emph {et~al.}}]{faurby2024purifying}%
  \BibitemOpen
  \bibfield  {author} {\bibinfo {author} {\bibfnamefont {C.~F.}\ \bibnamefont {Faurby}}, \bibinfo {author} {\bibfnamefont {L.}~\bibnamefont {Carosini}}, \bibinfo {author} {\bibfnamefont {H.}~\bibnamefont {Cao}}, \bibinfo {author} {\bibfnamefont {P.~I.}\ \bibnamefont {Sund}}, \bibinfo {author} {\bibfnamefont {L.~M.}\ \bibnamefont {Hansen}}, \bibinfo {author} {\bibfnamefont {F.}~\bibnamefont {Giorgino}}, \bibinfo {author} {\bibfnamefont {A.~B.}\ \bibnamefont {Villadsen}}, \bibinfo {author} {\bibfnamefont {S.~N.}\ \bibnamefont {Van Den~Hoven}}, \bibinfo {author} {\bibfnamefont {P.}~\bibnamefont {Lodahl}}, \bibinfo {author} {\bibfnamefont {S.}~\bibnamefont {Paesani}}, \emph {et~al.},\ }\bibfield  {title} {\bibinfo {title} {Purifying photon indistinguishability through quantum interference},\ }\href@noop {} {\bibfield  {journal} {\bibinfo  {journal} {Physical review letters}\ }\textbf {\bibinfo {volume} {133}},\ \bibinfo {pages} {033604} (\bibinfo {year} {2024})}\BibitemShut {NoStop}%
\bibitem [{\citenamefont {Bravyi}\ and\ \citenamefont {Kitaev}(2005)}]{bravyi2005universal}%
  \BibitemOpen
  \bibfield  {author} {\bibinfo {author} {\bibfnamefont {S.}~\bibnamefont {Bravyi}}\ and\ \bibinfo {author} {\bibfnamefont {A.}~\bibnamefont {Kitaev}},\ }\bibfield  {title} {\bibinfo {title} {Universal quantum computation with ideal clifford gates and noisy ancillas},\ }\href {https://doi.org/https://doi.org/10.1103/PhysRevA.71.022316} {\bibfield  {journal} {\bibinfo  {journal} {Physical Review A}\ }\textbf {\bibinfo {volume} {71}},\ \bibinfo {pages} {022316} (\bibinfo {year} {2005})}\BibitemShut {NoStop}%
\bibitem [{\citenamefont {Migdall}\ \emph {et~al.}(2002)\citenamefont {Migdall}, \citenamefont {Branning},\ and\ \citenamefont {Castelletto}}]{migdall2002tailoring}%
  \BibitemOpen
  \bibfield  {author} {\bibinfo {author} {\bibfnamefont {A.~L.}\ \bibnamefont {Migdall}}, \bibinfo {author} {\bibfnamefont {D.}~\bibnamefont {Branning}},\ and\ \bibinfo {author} {\bibfnamefont {S.}~\bibnamefont {Castelletto}},\ }\bibfield  {title} {\bibinfo {title} {Tailoring single-photon and multiphoton probabilities of a single-photon on-demand source},\ }\href {https://doi.org/https://doi.org/10.1103/PhysRevA.66.053805} {\bibfield  {journal} {\bibinfo  {journal} {Physical Review A}\ }\textbf {\bibinfo {volume} {66}},\ \bibinfo {pages} {053805} (\bibinfo {year} {2002})}\BibitemShut {NoStop}%
\bibitem [{\citenamefont {Shchesnovich}(2015)}]{shchesnovich2015partial}%
  \BibitemOpen
  \bibfield  {author} {\bibinfo {author} {\bibfnamefont {V.}~\bibnamefont {Shchesnovich}},\ }\bibfield  {title} {\bibinfo {title} {Partial indistinguishability theory for multiphoton experiments in multiport devices},\ }\href {https://doi.org/https://doi.org/10.1103/PhysRevA.91.013844} {\bibfield  {journal} {\bibinfo  {journal} {Physical Review A}\ }\textbf {\bibinfo {volume} {91}},\ \bibinfo {pages} {013844} (\bibinfo {year} {2015})}\BibitemShut {NoStop}%
\bibitem [{\citenamefont {Tichy}(2015)}]{tichy2015sampling}%
  \BibitemOpen
  \bibfield  {author} {\bibinfo {author} {\bibfnamefont {M.~C.}\ \bibnamefont {Tichy}},\ }\bibfield  {title} {\bibinfo {title} {Sampling of partially distinguishable bosons and the relation to the multidimensional permanent},\ }\href {https://doi.org/https://doi.org/10.1103/PhysRevA.91.022316} {\bibfield  {journal} {\bibinfo  {journal} {Physical Review A}\ }\textbf {\bibinfo {volume} {91}},\ \bibinfo {pages} {022316} (\bibinfo {year} {2015})}\BibitemShut {NoStop}%
\bibitem [{\citenamefont {Renema}\ \emph {et~al.}(2018)\citenamefont {Renema}, \citenamefont {Menssen}, \citenamefont {Clements}, \citenamefont {Triginer}, \citenamefont {Kolthammer},\ and\ \citenamefont {Walmsley}}]{renema2018efficient}%
  \BibitemOpen
  \bibfield  {author} {\bibinfo {author} {\bibfnamefont {J.~J.}\ \bibnamefont {Renema}}, \bibinfo {author} {\bibfnamefont {A.}~\bibnamefont {Menssen}}, \bibinfo {author} {\bibfnamefont {W.~R.}\ \bibnamefont {Clements}}, \bibinfo {author} {\bibfnamefont {G.}~\bibnamefont {Triginer}}, \bibinfo {author} {\bibfnamefont {W.~S.}\ \bibnamefont {Kolthammer}},\ and\ \bibinfo {author} {\bibfnamefont {I.~A.}\ \bibnamefont {Walmsley}},\ }\bibfield  {title} {\bibinfo {title} {Efficient classical algorithm for boson sampling with partially distinguishable photons},\ }\href {https://doi.org/https://doi.org/10.1103/PhysRevLett.120.220502} {\bibfield  {journal} {\bibinfo  {journal} {Physical Review Letters}\ }\textbf {\bibinfo {volume} {120}},\ \bibinfo {pages} {220502} (\bibinfo {year} {2018})}\BibitemShut {NoStop}%
\bibitem [{\citenamefont {Moylett}\ \emph {et~al.}(2019)\citenamefont {Moylett}, \citenamefont {Garc{\'\i}a-Patr{\'o}n}, \citenamefont {Renema},\ and\ \citenamefont {Turner}}]{moylett2019classically}%
  \BibitemOpen
  \bibfield  {author} {\bibinfo {author} {\bibfnamefont {A.~E.}\ \bibnamefont {Moylett}}, \bibinfo {author} {\bibfnamefont {R.}~\bibnamefont {Garc{\'\i}a-Patr{\'o}n}}, \bibinfo {author} {\bibfnamefont {J.~J.}\ \bibnamefont {Renema}},\ and\ \bibinfo {author} {\bibfnamefont {P.~S.}\ \bibnamefont {Turner}},\ }\bibfield  {title} {\bibinfo {title} {Classically simulating near-term partially-distinguishable and lossy boson sampling},\ }\href {https://doi.org/10.1088/2058-9565/ab5555} {\bibfield  {journal} {\bibinfo  {journal} {Quantum Science and Technology}\ }\textbf {\bibinfo {volume} {5}},\ \bibinfo {pages} {015001} (\bibinfo {year} {2019})}\BibitemShut {NoStop}%
\bibitem [{\citenamefont {Saied}\ \emph {et~al.}(2024{\natexlab{a}})\citenamefont {Saied}, \citenamefont {Marshall}, \citenamefont {Anand},\ and\ \citenamefont {Rieffel}}]{saied2024general}%
  \BibitemOpen
  \bibfield  {author} {\bibinfo {author} {\bibfnamefont {J.}~\bibnamefont {Saied}}, \bibinfo {author} {\bibfnamefont {J.}~\bibnamefont {Marshall}}, \bibinfo {author} {\bibfnamefont {N.}~\bibnamefont {Anand}},\ and\ \bibinfo {author} {\bibfnamefont {E.~G.}\ \bibnamefont {Rieffel}},\ }\bibfield  {title} {\bibinfo {title} {General protocols for the efficient distillation of indistinguishable photons},\ }\bibfield  {journal} {\bibinfo  {journal} {arXiv preprint arXiv:2404.14217}\ }\href {https://doi.org/https://doi.org/10.48550/arXiv.2404.14217} {https://doi.org/10.48550/arXiv.2404.14217} (\bibinfo {year} {2024}{\natexlab{a}})\BibitemShut {NoStop}%
\bibitem [{Note1()}]{Note1}%
  \BibitemOpen
  \bibinfo {note} {The \protect \textit {orthogonal bad-bit model} is equivalent to the \protect \textit {partial distinguishability model} where $x := 1 - \epsilon $ \cite {renema2018efficient}. The assumption of identical mutual partial distinguishability is demonstrated to be appropriate in multiphoton experiments with imperfect sources of sufficient quality \cite {renema2021sample}.}\BibitemShut {Stop}%
\bibitem [{Note2()}]{Note2}%
  \BibitemOpen
  \bibinfo {note} {We note that for a distilled photon, $\rho _{\protect \text {d}}$ is no longer a pure state but instead a maximally mixed state over the error modes of the initial resource photons. However, the uniqueness of these error modes is preserved in the mixed error state of the distilled photon. Consequently, the precise nature of $\rho _{\protect \text {d}}$ is irrelevant; for subsequent multi-photon interference, it is sufficient to treat $\rho _{\protect \text {d}}$ as equivalent to a pure, fully distinguishable photon state.}\BibitemShut {Stop}%
\bibitem [{\citenamefont {Arvind}\ \emph {et~al.}(1995)\citenamefont {Arvind}, \citenamefont {Dutta}, \citenamefont {Mukunda},\ and\ \citenamefont {Simon}}]{arvind1995real}%
  \BibitemOpen
  \bibfield  {author} {\bibinfo {author} {\bibnamefont {Arvind}}, \bibinfo {author} {\bibfnamefont {B.}~\bibnamefont {Dutta}}, \bibinfo {author} {\bibfnamefont {N.}~\bibnamefont {Mukunda}},\ and\ \bibinfo {author} {\bibfnamefont {R.}~\bibnamefont {Simon}},\ }\bibfield  {title} {\bibinfo {title} {The real symplectic groups in quantum mechanics and optics},\ }\href {https://doi.org/https://doi.org/10.1007/BF02848172} {\bibfield  {journal} {\bibinfo  {journal} {Pramana}\ }\textbf {\bibinfo {volume} {45}},\ \bibinfo {pages} {471} (\bibinfo {year} {1995})}\BibitemShut {NoStop}%
\bibitem [{\citenamefont {Tichy}\ \emph {et~al.}(2010)\citenamefont {Tichy}, \citenamefont {Tiersch}, \citenamefont {de~Melo}, \citenamefont {Mintert},\ and\ \citenamefont {Buchleitner}}]{tichy2010zero}%
  \BibitemOpen
  \bibfield  {author} {\bibinfo {author} {\bibfnamefont {M.~C.}\ \bibnamefont {Tichy}}, \bibinfo {author} {\bibfnamefont {M.}~\bibnamefont {Tiersch}}, \bibinfo {author} {\bibfnamefont {F.}~\bibnamefont {de~Melo}}, \bibinfo {author} {\bibfnamefont {F.}~\bibnamefont {Mintert}},\ and\ \bibinfo {author} {\bibfnamefont {A.}~\bibnamefont {Buchleitner}},\ }\bibfield  {title} {\bibinfo {title} {Zero-transmission law for multiport beam splitters},\ }\href {https://doi.org/https://doi.org/10.1103/PhysRevLett.104.220405} {\bibfield  {journal} {\bibinfo  {journal} {Physical Review Letters}\ }\textbf {\bibinfo {volume} {104}},\ \bibinfo {pages} {220405} (\bibinfo {year} {2010})}\BibitemShut {NoStop}%
\bibitem [{\citenamefont {Tichy}\ \emph {et~al.}(2014)\citenamefont {Tichy}, \citenamefont {Mayer}, \citenamefont {Buchleitner},\ and\ \citenamefont {M{\o}lmer}}]{tichy2014stringent}%
  \BibitemOpen
  \bibfield  {author} {\bibinfo {author} {\bibfnamefont {M.~C.}\ \bibnamefont {Tichy}}, \bibinfo {author} {\bibfnamefont {K.}~\bibnamefont {Mayer}}, \bibinfo {author} {\bibfnamefont {A.}~\bibnamefont {Buchleitner}},\ and\ \bibinfo {author} {\bibfnamefont {K.}~\bibnamefont {M{\o}lmer}},\ }\bibfield  {title} {\bibinfo {title} {Stringent and efficient assessment of boson-sampling devices},\ }\href {https://doi.org/https://doi.org/10.1103/PhysRevLett.113.020502} {\bibfield  {journal} {\bibinfo  {journal} {Physical Review Letters}\ }\textbf {\bibinfo {volume} {113}},\ \bibinfo {pages} {020502} (\bibinfo {year} {2014})}\BibitemShut {NoStop}%
\bibitem [{\citenamefont {Carolan}\ \emph {et~al.}(2015)\citenamefont {Carolan}, \citenamefont {Harrold}, \citenamefont {Sparrow}, \citenamefont {Mart{\'\i}n-L{\'o}pez}, \citenamefont {Russell}, \citenamefont {Silverstone}, \citenamefont {Shadbolt}, \citenamefont {Matsuda}, \citenamefont {Oguma}, \citenamefont {Itoh} \emph {et~al.}}]{carolan2015universal}%
  \BibitemOpen
  \bibfield  {author} {\bibinfo {author} {\bibfnamefont {J.}~\bibnamefont {Carolan}}, \bibinfo {author} {\bibfnamefont {C.}~\bibnamefont {Harrold}}, \bibinfo {author} {\bibfnamefont {C.}~\bibnamefont {Sparrow}}, \bibinfo {author} {\bibfnamefont {E.}~\bibnamefont {Mart{\'\i}n-L{\'o}pez}}, \bibinfo {author} {\bibfnamefont {N.~J.}\ \bibnamefont {Russell}}, \bibinfo {author} {\bibfnamefont {J.~W.}\ \bibnamefont {Silverstone}}, \bibinfo {author} {\bibfnamefont {P.~J.}\ \bibnamefont {Shadbolt}}, \bibinfo {author} {\bibfnamefont {N.}~\bibnamefont {Matsuda}}, \bibinfo {author} {\bibfnamefont {M.}~\bibnamefont {Oguma}}, \bibinfo {author} {\bibfnamefont {M.}~\bibnamefont {Itoh}}, \emph {et~al.},\ }\bibfield  {title} {\bibinfo {title} {Universal linear optics},\ }\href {https://doi.org/10.1126/science.aab3642} {\bibfield  {journal} {\bibinfo  {journal} {Science}\ }\textbf {\bibinfo {volume} {349}},\ \bibinfo {pages} {711} (\bibinfo {year} {2015})}\BibitemShut {NoStop}%
\bibitem [{\citenamefont {Crespi}\ \emph {et~al.}(2016)\citenamefont {Crespi}, \citenamefont {Osellame}, \citenamefont {Ramponi}, \citenamefont {Bentivegna}, \citenamefont {Flamini}, \citenamefont {Spagnolo}, \citenamefont {Viggianiello}, \citenamefont {Innocenti}, \citenamefont {Mataloni},\ and\ \citenamefont {Sciarrino}}]{crespi2016suppression}%
  \BibitemOpen
  \bibfield  {author} {\bibinfo {author} {\bibfnamefont {A.}~\bibnamefont {Crespi}}, \bibinfo {author} {\bibfnamefont {R.}~\bibnamefont {Osellame}}, \bibinfo {author} {\bibfnamefont {R.}~\bibnamefont {Ramponi}}, \bibinfo {author} {\bibfnamefont {M.}~\bibnamefont {Bentivegna}}, \bibinfo {author} {\bibfnamefont {F.}~\bibnamefont {Flamini}}, \bibinfo {author} {\bibfnamefont {N.}~\bibnamefont {Spagnolo}}, \bibinfo {author} {\bibfnamefont {N.}~\bibnamefont {Viggianiello}}, \bibinfo {author} {\bibfnamefont {L.}~\bibnamefont {Innocenti}}, \bibinfo {author} {\bibfnamefont {P.}~\bibnamefont {Mataloni}},\ and\ \bibinfo {author} {\bibfnamefont {F.}~\bibnamefont {Sciarrino}},\ }\bibfield  {title} {\bibinfo {title} {Suppression law of quantum states in a 3d photonic fast fourier transform chip},\ }\href {https://doi.org/https://doi.org/10.1038/ncomms10469} {\bibfield  {journal} {\bibinfo  {journal} {Nature Communications}\ }\textbf {\bibinfo {volume} {7}},\ \bibinfo {pages} {10469} (\bibinfo {year} {2016})}\BibitemShut
  {NoStop}%
\bibitem [{\citenamefont {Stanisic}\ and\ \citenamefont {Turner}(2018)}]{stanisic2018discriminating}%
  \BibitemOpen
  \bibfield  {author} {\bibinfo {author} {\bibfnamefont {S.}~\bibnamefont {Stanisic}}\ and\ \bibinfo {author} {\bibfnamefont {P.~S.}\ \bibnamefont {Turner}},\ }\bibfield  {title} {\bibinfo {title} {Discriminating distinguishability},\ }\href {https://doi.org/https://doi.org/10.1103/PhysRevA.98.043839} {\bibfield  {journal} {\bibinfo  {journal} {Physical Review A}\ }\textbf {\bibinfo {volume} {98}},\ \bibinfo {pages} {043839} (\bibinfo {year} {2018})}\BibitemShut {NoStop}%
\bibitem [{\citenamefont {Dittel}\ \emph {et~al.}(2018)\citenamefont {Dittel}, \citenamefont {Dufour}, \citenamefont {Walschaers}, \citenamefont {Weihs}, \citenamefont {Buchleitner},\ and\ \citenamefont {Keil}}]{dittel2018totally}%
  \BibitemOpen
  \bibfield  {author} {\bibinfo {author} {\bibfnamefont {C.}~\bibnamefont {Dittel}}, \bibinfo {author} {\bibfnamefont {G.}~\bibnamefont {Dufour}}, \bibinfo {author} {\bibfnamefont {M.}~\bibnamefont {Walschaers}}, \bibinfo {author} {\bibfnamefont {G.}~\bibnamefont {Weihs}}, \bibinfo {author} {\bibfnamefont {A.}~\bibnamefont {Buchleitner}},\ and\ \bibinfo {author} {\bibfnamefont {R.}~\bibnamefont {Keil}},\ }\bibfield  {title} {\bibinfo {title} {Totally destructive many-particle interference},\ }\href {https://doi.org/https://doi.org/10.1103/PhysRevLett.120.240404} {\bibfield  {journal} {\bibinfo  {journal} {Physical Review Letters}\ }\textbf {\bibinfo {volume} {120}},\ \bibinfo {pages} {240404} (\bibinfo {year} {2018})}\BibitemShut {NoStop}%
\bibitem [{\citenamefont {Brod}\ \emph {et~al.}(2019)\citenamefont {Brod}, \citenamefont {Galv{\~a}o}, \citenamefont {Viggianiello}, \citenamefont {Flamini}, \citenamefont {Spagnolo},\ and\ \citenamefont {Sciarrino}}]{brod2019witnessing}%
  \BibitemOpen
  \bibfield  {author} {\bibinfo {author} {\bibfnamefont {D.~J.}\ \bibnamefont {Brod}}, \bibinfo {author} {\bibfnamefont {E.~F.}\ \bibnamefont {Galv{\~a}o}}, \bibinfo {author} {\bibfnamefont {N.}~\bibnamefont {Viggianiello}}, \bibinfo {author} {\bibfnamefont {F.}~\bibnamefont {Flamini}}, \bibinfo {author} {\bibfnamefont {N.}~\bibnamefont {Spagnolo}},\ and\ \bibinfo {author} {\bibfnamefont {F.}~\bibnamefont {Sciarrino}},\ }\bibfield  {title} {\bibinfo {title} {Witnessing genuine multiphoton indistinguishability},\ }\href {https://doi.org/https://doi.org/10.1103/PhysRevLett.122.063602} {\bibfield  {journal} {\bibinfo  {journal} {Physical Review Letters}\ }\textbf {\bibinfo {volume} {122}},\ \bibinfo {pages} {063602} (\bibinfo {year} {2019})}\BibitemShut {NoStop}%
\bibitem [{\citenamefont {Somhorst}\ \emph {et~al.}(2023)\citenamefont {Somhorst}, \citenamefont {van~der Meer}, \citenamefont {Correa~Anguita}, \citenamefont {Schadow}, \citenamefont {Snijders}, \citenamefont {de~Goede}, \citenamefont {Kassenberg}, \citenamefont {Venderbosch}, \citenamefont {Taballione}, \citenamefont {Epping} \emph {et~al.}}]{somhorst2023quantum}%
  \BibitemOpen
  \bibfield  {author} {\bibinfo {author} {\bibfnamefont {F.~H.~B.}\ \bibnamefont {Somhorst}}, \bibinfo {author} {\bibfnamefont {R.}~\bibnamefont {van~der Meer}}, \bibinfo {author} {\bibfnamefont {M.}~\bibnamefont {Correa~Anguita}}, \bibinfo {author} {\bibfnamefont {R.}~\bibnamefont {Schadow}}, \bibinfo {author} {\bibfnamefont {H.~J.}\ \bibnamefont {Snijders}}, \bibinfo {author} {\bibfnamefont {M.}~\bibnamefont {de~Goede}}, \bibinfo {author} {\bibfnamefont {B.}~\bibnamefont {Kassenberg}}, \bibinfo {author} {\bibfnamefont {P.}~\bibnamefont {Venderbosch}}, \bibinfo {author} {\bibfnamefont {C.}~\bibnamefont {Taballione}}, \bibinfo {author} {\bibfnamefont {J.}~\bibnamefont {Epping}}, \emph {et~al.},\ }\bibfield  {title} {\bibinfo {title} {Quantum simulation of thermodynamics in an integrated quantum photonic processor},\ }\href {https://doi.org/https://doi.org/10.1038/s41467-023-38413-9} {\bibfield  {journal} {\bibinfo  {journal} {Nature Communications}\ }\textbf {\bibinfo {volume} {14}},\ \bibinfo {pages} {3895}
  (\bibinfo {year} {2023})}\BibitemShut {NoStop}%
\bibitem [{Note3()}]{Note3}%
  \BibitemOpen
  \bibinfo {note} {For example, the mode-assignment list $(1,2,3,4,5)$ corresponds to the equivalent non-collision state $\mathinner {|{11111}\rangle }$.}\BibitemShut {Stop}%
\bibitem [{Note4()}]{Note4}%
  \BibitemOpen
  \bibinfo {note} {However, a classification as \protect \textit {allowed outcome} is a necessary but not sufficient condition for a classification as \protect \textit {non-suppressed outcome}. Nevertheless, most allowed outcomes constitute non-suppressed outcomes \cite {tichy2010zero}.}\BibitemShut {Stop}%
\bibitem [{\citenamefont {Somhorst}\ \emph {et~al.}(2024)\citenamefont {Somhorst}, \citenamefont {Sauër}, \citenamefont {van~den Hoven},\ and\ \citenamefont {Renema}}]{dataSet}%
  \BibitemOpen
  \bibfield  {author} {\bibinfo {author} {\bibfnamefont {F.~H.~B.}\ \bibnamefont {Somhorst}}, \bibinfo {author} {\bibfnamefont {B.}~\bibnamefont {Sauër}}, \bibinfo {author} {\bibfnamefont {S.}~\bibnamefont {van~den Hoven}},\ and\ \bibinfo {author} {\bibfnamefont {J.}~\bibnamefont {Renema}},\ }\bibfield  {title} {\bibinfo {title} {"data underlying the manuscript: "photon distillation schemes with reduced resource costs based on multiphoton fourier interference"},\ }\bibfield  {journal} {\bibinfo  {journal} {4TU.ResearchData}\ }\href {https://doi.org/10.4121/c0c83c18-f38b-46b8-b4f4-67964cb611d0} {10.4121/c0c83c18-f38b-46b8-b4f4-67964cb611d0} (\bibinfo {year} {2024})\BibitemShut {NoStop}%
\bibitem [{\citenamefont {Yung}\ \emph {et~al.}(2019)\citenamefont {Yung}, \citenamefont {Gao},\ and\ \citenamefont {Huh}}]{Joonsuk2019bunching}%
  \BibitemOpen
  \bibfield  {author} {\bibinfo {author} {\bibfnamefont {M.-H.}\ \bibnamefont {Yung}}, \bibinfo {author} {\bibfnamefont {X.}~\bibnamefont {Gao}},\ and\ \bibinfo {author} {\bibfnamefont {J.}~\bibnamefont {Huh}},\ }\bibfield  {title} {\bibinfo {title} {Universal bound on sampling bosons in linear optics and its computational implications},\ }\href {https://doi.org/https://doi.org/10.1093/nsr/nwz048} {\bibfield  {journal} {\bibinfo  {journal} {National Science Review}\ }\textbf {\bibinfo {volume} {6}},\ \bibinfo {pages} {719} (\bibinfo {year} {2019})}\BibitemShut {NoStop}%
\bibitem [{\citenamefont {Seron}\ \emph {et~al.}(2023)\citenamefont {Seron}, \citenamefont {Novo},\ and\ \citenamefont {Cerf}}]{seron2023boson}%
  \BibitemOpen
  \bibfield  {author} {\bibinfo {author} {\bibfnamefont {B.}~\bibnamefont {Seron}}, \bibinfo {author} {\bibfnamefont {L.}~\bibnamefont {Novo}},\ and\ \bibinfo {author} {\bibfnamefont {N.~J.}\ \bibnamefont {Cerf}},\ }\bibfield  {title} {\bibinfo {title} {Boson bunching is not maximized by indistinguishable particles},\ }\href@noop {} {\bibfield  {journal} {\bibinfo  {journal} {Nature Photonics}\ }\textbf {\bibinfo {volume} {17}},\ \bibinfo {pages} {702} (\bibinfo {year} {2023})}\BibitemShut {NoStop}%
\bibitem [{\citenamefont {Clifford}\ and\ \citenamefont {Clifford}(2018)}]{clifford2018classical}%
  \BibitemOpen
  \bibfield  {author} {\bibinfo {author} {\bibfnamefont {P.}~\bibnamefont {Clifford}}\ and\ \bibinfo {author} {\bibfnamefont {R.}~\bibnamefont {Clifford}},\ }\bibfield  {title} {\bibinfo {title} {The classical complexity of boson sampling},\ }in\ \href {https://doi.org/https://doi.org/10.48550/arXiv.1706.01260} {\emph {\bibinfo {booktitle} {Proceedings of the Twenty-Ninth Annual ACM-SIAM Symposium on Discrete Algorithms}}}\ (\bibinfo {organization} {SIAM},\ \bibinfo {year} {2018})\ pp.\ \bibinfo {pages} {146--155}\BibitemShut {NoStop}%
\bibitem [{\citenamefont {Saied}\ and\ \citenamefont {Marshall}()}]{privateCom}%
  \BibitemOpen
  \bibfield  {author} {\bibinfo {author} {\bibfnamefont {J.}~\bibnamefont {Saied}}\ and\ \bibinfo {author} {\bibfnamefont {J.}~\bibnamefont {Marshall}},\ }\href@noop {} {}\bibinfo {howpublished} {private communication}\BibitemShut {NoStop}%
\bibitem [{Note5()}]{Note5}%
  \BibitemOpen
  \bibinfo {note} {Pre-computed data for $N = 3$ to $N = 10$ can be found in \cite {dataSet}}\BibitemShut {NoStop}%
\bibitem [{\citenamefont {Aharonov}\ and\ \citenamefont {Ben-Or}(1997)}]{aharonov1997fault}%
  \BibitemOpen
  \bibfield  {author} {\bibinfo {author} {\bibfnamefont {D.}~\bibnamefont {Aharonov}}\ and\ \bibinfo {author} {\bibfnamefont {M.}~\bibnamefont {Ben-Or}},\ }\bibfield  {title} {\bibinfo {title} {Fault-tolerant quantum computation with constant error},\ }in\ \href@noop {} {\emph {\bibinfo {booktitle} {Proceedings of the twenty-ninth annual ACM symposium on Theory of computing}}}\ (\bibinfo {year} {1997})\ pp.\ \bibinfo {pages} {176--188}\BibitemShut {NoStop}%
\bibitem [{\citenamefont {Alexander}\ \emph {et~al.}(2024)\citenamefont {Alexander}, \citenamefont {Bahgat}, \citenamefont {Benyamini}, \citenamefont {Black}, \citenamefont {Bonneau}, \citenamefont {Burgos}, \citenamefont {Burridge}, \citenamefont {Campbell}, \citenamefont {Catalano}, \citenamefont {Ceballos} \emph {et~al.}}]{alexander2024manufacturable}%
  \BibitemOpen
  \bibfield  {author} {\bibinfo {author} {\bibfnamefont {K.}~\bibnamefont {Alexander}}, \bibinfo {author} {\bibfnamefont {A.}~\bibnamefont {Bahgat}}, \bibinfo {author} {\bibfnamefont {A.}~\bibnamefont {Benyamini}}, \bibinfo {author} {\bibfnamefont {D.}~\bibnamefont {Black}}, \bibinfo {author} {\bibfnamefont {D.}~\bibnamefont {Bonneau}}, \bibinfo {author} {\bibfnamefont {S.}~\bibnamefont {Burgos}}, \bibinfo {author} {\bibfnamefont {B.}~\bibnamefont {Burridge}}, \bibinfo {author} {\bibfnamefont {G.}~\bibnamefont {Campbell}}, \bibinfo {author} {\bibfnamefont {G.}~\bibnamefont {Catalano}}, \bibinfo {author} {\bibfnamefont {A.}~\bibnamefont {Ceballos}}, \emph {et~al.},\ }\bibfield  {title} {\bibinfo {title} {A manufacturable platform for photonic quantum computing},\ }\href@noop {} {\bibfield  {journal} {\bibinfo  {journal} {arXiv preprint arXiv:2404.17570}\ } (\bibinfo {year} {2024})}\BibitemShut {NoStop}%
\bibitem [{\citenamefont {Rohde}\ and\ \citenamefont {Ralph}(2006)}]{rohde2006error}%
  \BibitemOpen
  \bibfield  {author} {\bibinfo {author} {\bibfnamefont {P.~P.}\ \bibnamefont {Rohde}}\ and\ \bibinfo {author} {\bibfnamefont {T.~C.}\ \bibnamefont {Ralph}},\ }\bibfield  {title} {\bibinfo {title} {Error models for mode mismatch in linear optics quantum computing},\ }\href@noop {} {\bibfield  {journal} {\bibinfo  {journal} {Physical Review A—Atomic, Molecular, and Optical Physics}\ }\textbf {\bibinfo {volume} {73}},\ \bibinfo {pages} {062312} (\bibinfo {year} {2006})}\BibitemShut {NoStop}%
\bibitem [{\citenamefont {Saied}\ \emph {et~al.}(2024{\natexlab{b}})\citenamefont {Saied}, \citenamefont {Marshall}, \citenamefont {Anand}, \citenamefont {Grabbe},\ and\ \citenamefont {Rieffel}}]{saied2024advancing}%
  \BibitemOpen
  \bibfield  {author} {\bibinfo {author} {\bibfnamefont {J.}~\bibnamefont {Saied}}, \bibinfo {author} {\bibfnamefont {J.}~\bibnamefont {Marshall}}, \bibinfo {author} {\bibfnamefont {N.}~\bibnamefont {Anand}}, \bibinfo {author} {\bibfnamefont {S.}~\bibnamefont {Grabbe}},\ and\ \bibinfo {author} {\bibfnamefont {E.~G.}\ \bibnamefont {Rieffel}},\ }\bibfield  {title} {\bibinfo {title} {Advancing quantum networking: some tools and protocols for ideal and noisy photonic systems},\ }in\ \href@noop {} {\emph {\bibinfo {booktitle} {Quantum Computing, Communication, and Simulation IV}}},\ Vol.\ \bibinfo {volume} {12911}\ (\bibinfo {organization} {SPIE},\ \bibinfo {year} {2024})\ pp.\ \bibinfo {pages} {37--65}\BibitemShut {NoStop}%
\bibitem [{\citenamefont {Fowler}\ \emph {et~al.}(2012)\citenamefont {Fowler}, \citenamefont {Mariantoni}, \citenamefont {Martinis},\ and\ \citenamefont {Cleland}}]{fowler2012surface}%
  \BibitemOpen
  \bibfield  {author} {\bibinfo {author} {\bibfnamefont {A.~G.}\ \bibnamefont {Fowler}}, \bibinfo {author} {\bibfnamefont {M.}~\bibnamefont {Mariantoni}}, \bibinfo {author} {\bibfnamefont {J.~M.}\ \bibnamefont {Martinis}},\ and\ \bibinfo {author} {\bibfnamefont {A.~N.}\ \bibnamefont {Cleland}},\ }\bibfield  {title} {\bibinfo {title} {Surface codes: Towards practical large-scale quantum computation},\ }\href@noop {} {\bibfield  {journal} {\bibinfo  {journal} {Physical Review A}\ }\textbf {\bibinfo {volume} {86}},\ \bibinfo {pages} {032324} (\bibinfo {year} {2012})}\BibitemShut {NoStop}%
\bibitem [{Note6()}]{Note6}%
  \BibitemOpen
  \bibinfo {note} {This is a conservative model: if $p_{\protect \text {error}} = \protect \mathcal {O}(\epsilon ^k)$ for integer $k \geq 2$, the impact of photon distillation will be even greater, as the factor $1/N$ in Eq. \ref {eq:inequalityQEC} will be replaced by $1/N^k$.}\BibitemShut {Stop}%
\bibitem [{\citenamefont {Harris}\ \emph {et~al.}(2017)\citenamefont {Harris}, \citenamefont {Steinbrecher}, \citenamefont {Prabhu}, \citenamefont {Lahini}, \citenamefont {Mower}, \citenamefont {Bunandar}, \citenamefont {Chen}, \citenamefont {Wong}, \citenamefont {Baehr-Jones}, \citenamefont {Hochberg} \emph {et~al.}}]{harris2017quantum}%
  \BibitemOpen
  \bibfield  {author} {\bibinfo {author} {\bibfnamefont {N.~C.}\ \bibnamefont {Harris}}, \bibinfo {author} {\bibfnamefont {G.~R.}\ \bibnamefont {Steinbrecher}}, \bibinfo {author} {\bibfnamefont {M.}~\bibnamefont {Prabhu}}, \bibinfo {author} {\bibfnamefont {Y.}~\bibnamefont {Lahini}}, \bibinfo {author} {\bibfnamefont {J.}~\bibnamefont {Mower}}, \bibinfo {author} {\bibfnamefont {D.}~\bibnamefont {Bunandar}}, \bibinfo {author} {\bibfnamefont {C.}~\bibnamefont {Chen}}, \bibinfo {author} {\bibfnamefont {F.~N.}\ \bibnamefont {Wong}}, \bibinfo {author} {\bibfnamefont {T.}~\bibnamefont {Baehr-Jones}}, \bibinfo {author} {\bibfnamefont {M.}~\bibnamefont {Hochberg}}, \emph {et~al.},\ }\bibfield  {title} {\bibinfo {title} {Quantum transport simulations in a programmable nanophotonic processor},\ }\href {https://doi.org/https://doi.org/10.1038/nphoton.2017.95} {\bibfield  {journal} {\bibinfo  {journal} {Nature Photonics}\ }\textbf {\bibinfo {volume} {11}},\ \bibinfo {pages} {447} (\bibinfo {year} {2017})}\BibitemShut
  {NoStop}%
\bibitem [{\citenamefont {Sparrow}\ \emph {et~al.}(2018)\citenamefont {Sparrow}, \citenamefont {Mart{\'\i}n-L{\'o}pez}, \citenamefont {Maraviglia}, \citenamefont {Neville}, \citenamefont {Harrold}, \citenamefont {Carolan}, \citenamefont {Joglekar}, \citenamefont {Hashimoto}, \citenamefont {Matsuda}, \citenamefont {O’Brien} \emph {et~al.}}]{sparrow2018simulating}%
  \BibitemOpen
  \bibfield  {author} {\bibinfo {author} {\bibfnamefont {C.}~\bibnamefont {Sparrow}}, \bibinfo {author} {\bibfnamefont {E.}~\bibnamefont {Mart{\'\i}n-L{\'o}pez}}, \bibinfo {author} {\bibfnamefont {N.}~\bibnamefont {Maraviglia}}, \bibinfo {author} {\bibfnamefont {A.}~\bibnamefont {Neville}}, \bibinfo {author} {\bibfnamefont {C.}~\bibnamefont {Harrold}}, \bibinfo {author} {\bibfnamefont {J.}~\bibnamefont {Carolan}}, \bibinfo {author} {\bibfnamefont {Y.~N.}\ \bibnamefont {Joglekar}}, \bibinfo {author} {\bibfnamefont {T.}~\bibnamefont {Hashimoto}}, \bibinfo {author} {\bibfnamefont {N.}~\bibnamefont {Matsuda}}, \bibinfo {author} {\bibfnamefont {J.~L.}\ \bibnamefont {O’Brien}}, \emph {et~al.},\ }\bibfield  {title} {\bibinfo {title} {Simulating the vibrational quantum dynamics of molecules using photonics},\ }\href {https://doi.org/https://doi.org/10.1038/s41586-018-0152-9} {\bibfield  {journal} {\bibinfo  {journal} {Nature}\ }\textbf {\bibinfo {volume} {557}},\ \bibinfo {pages} {660} (\bibinfo {year}
  {2018})}\BibitemShut {NoStop}%
\bibitem [{\citenamefont {Wang}\ \emph {et~al.}(2020)\citenamefont {Wang}, \citenamefont {Sciarrino}, \citenamefont {Laing},\ and\ \citenamefont {Thompson}}]{wang2020integrated}%
  \BibitemOpen
  \bibfield  {author} {\bibinfo {author} {\bibfnamefont {J.}~\bibnamefont {Wang}}, \bibinfo {author} {\bibfnamefont {F.}~\bibnamefont {Sciarrino}}, \bibinfo {author} {\bibfnamefont {A.}~\bibnamefont {Laing}},\ and\ \bibinfo {author} {\bibfnamefont {M.~G.}\ \bibnamefont {Thompson}},\ }\bibfield  {title} {\bibinfo {title} {Integrated photonic quantum technologies},\ }\href {https://doi.org/https://doi.org/10.1038/s41566-019-0532-1} {\bibfield  {journal} {\bibinfo  {journal} {Nature Photonics}\ }\textbf {\bibinfo {volume} {14}},\ \bibinfo {pages} {273} (\bibinfo {year} {2020})}\BibitemShut {NoStop}%
\bibitem [{\citenamefont {Taballione}\ \emph {et~al.}(2021)\citenamefont {Taballione}, \citenamefont {van~der Meer}, \citenamefont {Snijders}, \citenamefont {Hooijschuur}, \citenamefont {Epping}, \citenamefont {de~Goede}, \citenamefont {Kassenberg}, \citenamefont {Venderbosch}, \citenamefont {Toebes}, \citenamefont {van~den Vlekkert} \emph {et~al.}}]{taballione2021universal}%
  \BibitemOpen
  \bibfield  {author} {\bibinfo {author} {\bibfnamefont {C.}~\bibnamefont {Taballione}}, \bibinfo {author} {\bibfnamefont {R.}~\bibnamefont {van~der Meer}}, \bibinfo {author} {\bibfnamefont {H.~J.}\ \bibnamefont {Snijders}}, \bibinfo {author} {\bibfnamefont {P.}~\bibnamefont {Hooijschuur}}, \bibinfo {author} {\bibfnamefont {J.~P.}\ \bibnamefont {Epping}}, \bibinfo {author} {\bibfnamefont {M.}~\bibnamefont {de~Goede}}, \bibinfo {author} {\bibfnamefont {B.}~\bibnamefont {Kassenberg}}, \bibinfo {author} {\bibfnamefont {P.}~\bibnamefont {Venderbosch}}, \bibinfo {author} {\bibfnamefont {C.}~\bibnamefont {Toebes}}, \bibinfo {author} {\bibfnamefont {H.}~\bibnamefont {van~den Vlekkert}}, \emph {et~al.},\ }\bibfield  {title} {\bibinfo {title} {A universal fully reconfigurable 12-mode quantum photonic processor},\ }\href {https://doi.org/https://doi.org/10.1088/2633-4356/ac168c} {\bibfield  {journal} {\bibinfo  {journal} {Materials for Quantum Technology}\ }\textbf {\bibinfo {volume} {1}},\ \bibinfo {pages} {035002}
  (\bibinfo {year} {2021})}\BibitemShut {NoStop}%
\bibitem [{\citenamefont {Taballione}\ \emph {et~al.}(2023)\citenamefont {Taballione}, \citenamefont {Anguita}, \citenamefont {de~Goede}, \citenamefont {Venderbosch}, \citenamefont {Kassenberg}, \citenamefont {Snijders}, \citenamefont {Kannan}, \citenamefont {Vleeshouwers}, \citenamefont {Smith}, \citenamefont {Epping} \emph {et~al.}}]{taballione202320}%
  \BibitemOpen
  \bibfield  {author} {\bibinfo {author} {\bibfnamefont {C.}~\bibnamefont {Taballione}}, \bibinfo {author} {\bibfnamefont {M.~C.}\ \bibnamefont {Anguita}}, \bibinfo {author} {\bibfnamefont {M.}~\bibnamefont {de~Goede}}, \bibinfo {author} {\bibfnamefont {P.}~\bibnamefont {Venderbosch}}, \bibinfo {author} {\bibfnamefont {B.}~\bibnamefont {Kassenberg}}, \bibinfo {author} {\bibfnamefont {H.}~\bibnamefont {Snijders}}, \bibinfo {author} {\bibfnamefont {N.}~\bibnamefont {Kannan}}, \bibinfo {author} {\bibfnamefont {W.~L.}\ \bibnamefont {Vleeshouwers}}, \bibinfo {author} {\bibfnamefont {D.}~\bibnamefont {Smith}}, \bibinfo {author} {\bibfnamefont {J.~P.}\ \bibnamefont {Epping}}, \emph {et~al.},\ }\bibfield  {title} {\bibinfo {title} {20-mode universal quantum photonic processor},\ }\href {https://doi.org/https://doi.org/10.22331/q-2023-08-01-1071} {\bibfield  {journal} {\bibinfo  {journal} {Quantum}\ }\textbf {\bibinfo {volume} {7}},\ \bibinfo {pages} {1071} (\bibinfo {year} {2023})}\BibitemShut {NoStop}%
\bibitem [{\citenamefont {Tomm}\ \emph {et~al.}(2021)\citenamefont {Tomm}, \citenamefont {Javadi}, \citenamefont {Antoniadis}, \citenamefont {Najer}, \citenamefont {L{\"o}bl}, \citenamefont {Korsch}, \citenamefont {Schott}, \citenamefont {Valentin}, \citenamefont {Wieck}, \citenamefont {Ludwig} \emph {et~al.}}]{tomm2021bright}%
  \BibitemOpen
  \bibfield  {author} {\bibinfo {author} {\bibfnamefont {N.}~\bibnamefont {Tomm}}, \bibinfo {author} {\bibfnamefont {A.}~\bibnamefont {Javadi}}, \bibinfo {author} {\bibfnamefont {N.~O.}\ \bibnamefont {Antoniadis}}, \bibinfo {author} {\bibfnamefont {D.}~\bibnamefont {Najer}}, \bibinfo {author} {\bibfnamefont {M.~C.}\ \bibnamefont {L{\"o}bl}}, \bibinfo {author} {\bibfnamefont {A.~R.}\ \bibnamefont {Korsch}}, \bibinfo {author} {\bibfnamefont {R.}~\bibnamefont {Schott}}, \bibinfo {author} {\bibfnamefont {S.~R.}\ \bibnamefont {Valentin}}, \bibinfo {author} {\bibfnamefont {A.~D.}\ \bibnamefont {Wieck}}, \bibinfo {author} {\bibfnamefont {A.}~\bibnamefont {Ludwig}}, \emph {et~al.},\ }\bibfield  {title} {\bibinfo {title} {A bright and fast source of coherent single photons},\ }\href {https://doi.org/https://doi.org/10.1038/s41565-020-00831-x} {\bibfield  {journal} {\bibinfo  {journal} {Nature Nanotechnology}\ }\textbf {\bibinfo {volume} {16}},\ \bibinfo {pages} {399} (\bibinfo {year} {2021})}\BibitemShut {NoStop}%
\bibitem [{\citenamefont {Ollivier}\ \emph {et~al.}(2021)\citenamefont {Ollivier}, \citenamefont {Thomas}, \citenamefont {Wein}, \citenamefont {de~Buy~Wenniger}, \citenamefont {Coste}, \citenamefont {Loredo}, \citenamefont {Somaschi}, \citenamefont {Harouri}, \citenamefont {Lemaitre}, \citenamefont {Sagnes} \emph {et~al.}}]{ollivier2021hong}%
  \BibitemOpen
  \bibfield  {author} {\bibinfo {author} {\bibfnamefont {H.}~\bibnamefont {Ollivier}}, \bibinfo {author} {\bibfnamefont {S.}~\bibnamefont {Thomas}}, \bibinfo {author} {\bibfnamefont {S.}~\bibnamefont {Wein}}, \bibinfo {author} {\bibfnamefont {I.~M.}\ \bibnamefont {de~Buy~Wenniger}}, \bibinfo {author} {\bibfnamefont {N.}~\bibnamefont {Coste}}, \bibinfo {author} {\bibfnamefont {J.}~\bibnamefont {Loredo}}, \bibinfo {author} {\bibfnamefont {N.}~\bibnamefont {Somaschi}}, \bibinfo {author} {\bibfnamefont {A.}~\bibnamefont {Harouri}}, \bibinfo {author} {\bibfnamefont {A.}~\bibnamefont {Lemaitre}}, \bibinfo {author} {\bibfnamefont {I.}~\bibnamefont {Sagnes}}, \emph {et~al.},\ }\bibfield  {title} {\bibinfo {title} {Hong-ou-mandel interference with imperfect single photon sources},\ }\href {https://doi.org/10.1103/PhysRevLett.126.063602} {\bibfield  {journal} {\bibinfo  {journal} {Physical Review Letters}\ }\textbf {\bibinfo {volume} {126}},\ \bibinfo {pages} {063602} (\bibinfo {year} {2021})}\BibitemShut {NoStop}%
\bibitem [{\citenamefont {Becher}\ \emph {et~al.}(2023)\citenamefont {Becher}, \citenamefont {Gao}, \citenamefont {Kar}, \citenamefont {Marciniak}, \citenamefont {Monz}, \citenamefont {Bartholomew}, \citenamefont {Goldner}, \citenamefont {Loh}, \citenamefont {Marcellina}, \citenamefont {Goh} \emph {et~al.}}]{becher20232023}%
  \BibitemOpen
  \bibfield  {author} {\bibinfo {author} {\bibfnamefont {C.}~\bibnamefont {Becher}}, \bibinfo {author} {\bibfnamefont {W.}~\bibnamefont {Gao}}, \bibinfo {author} {\bibfnamefont {S.}~\bibnamefont {Kar}}, \bibinfo {author} {\bibfnamefont {C.~D.}\ \bibnamefont {Marciniak}}, \bibinfo {author} {\bibfnamefont {T.}~\bibnamefont {Monz}}, \bibinfo {author} {\bibfnamefont {J.~G.}\ \bibnamefont {Bartholomew}}, \bibinfo {author} {\bibfnamefont {P.}~\bibnamefont {Goldner}}, \bibinfo {author} {\bibfnamefont {H.}~\bibnamefont {Loh}}, \bibinfo {author} {\bibfnamefont {E.}~\bibnamefont {Marcellina}}, \bibinfo {author} {\bibfnamefont {K.~E.~J.}\ \bibnamefont {Goh}}, \emph {et~al.},\ }\bibfield  {title} {\bibinfo {title} {2023 roadmap for materials for quantum technologies},\ }\href {https://doi.org/10.1088/2633-4356/aca3f2} {\bibfield  {journal} {\bibinfo  {journal} {Materials for Quantum Technology}\ }\textbf {\bibinfo {volume} {3}},\ \bibinfo {pages} {012501} (\bibinfo {year} {2023})}\BibitemShut {NoStop}%
\bibitem [{\citenamefont {Ding}\ \emph {et~al.}(2023)\citenamefont {Ding}, \citenamefont {Guo}, \citenamefont {Xu}, \citenamefont {Liu}, \citenamefont {Zou}, \citenamefont {Zhao}, \citenamefont {Ge}, \citenamefont {Zhang}, \citenamefont {Liu}, \citenamefont {Chen} \emph {et~al.}}]{ding2023high}%
  \BibitemOpen
  \bibfield  {author} {\bibinfo {author} {\bibfnamefont {X.}~\bibnamefont {Ding}}, \bibinfo {author} {\bibfnamefont {Y.-P.}\ \bibnamefont {Guo}}, \bibinfo {author} {\bibfnamefont {M.-C.}\ \bibnamefont {Xu}}, \bibinfo {author} {\bibfnamefont {R.-Z.}\ \bibnamefont {Liu}}, \bibinfo {author} {\bibfnamefont {G.-Y.}\ \bibnamefont {Zou}}, \bibinfo {author} {\bibfnamefont {J.-Y.}\ \bibnamefont {Zhao}}, \bibinfo {author} {\bibfnamefont {Z.-X.}\ \bibnamefont {Ge}}, \bibinfo {author} {\bibfnamefont {Q.-H.}\ \bibnamefont {Zhang}}, \bibinfo {author} {\bibfnamefont {H.-L.}\ \bibnamefont {Liu}}, \bibinfo {author} {\bibfnamefont {M.-C.}\ \bibnamefont {Chen}}, \emph {et~al.},\ }\bibfield  {title} {\bibinfo {title} {High-efficiency single-photon source above the loss-tolerant threshold for efficient linear optical quantum computing},\ }\bibfield  {journal} {\bibinfo  {journal} {arXiv preprint arXiv:2311.08347}\ }\href {https://doi.org/https://doi.org/10.48550/arXiv.2311.08347} {https://doi.org/10.48550/arXiv.2311.08347}
  (\bibinfo {year} {2023})\BibitemShut {NoStop}%
\bibitem [{\citenamefont {Renema}\ \emph {et~al.}(2021)\citenamefont {Renema}, \citenamefont {Wang}, \citenamefont {Qin}, \citenamefont {You}, \citenamefont {Lu},\ and\ \citenamefont {Pan}}]{renema2021sample}%
  \BibitemOpen
  \bibfield  {author} {\bibinfo {author} {\bibfnamefont {J.~J.}\ \bibnamefont {Renema}}, \bibinfo {author} {\bibfnamefont {H.}~\bibnamefont {Wang}}, \bibinfo {author} {\bibfnamefont {J.}~\bibnamefont {Qin}}, \bibinfo {author} {\bibfnamefont {X.}~\bibnamefont {You}}, \bibinfo {author} {\bibfnamefont {C.}~\bibnamefont {Lu}},\ and\ \bibinfo {author} {\bibfnamefont {J.}~\bibnamefont {Pan}},\ }\bibfield  {title} {\bibinfo {title} {Sample-efficient benchmarking of multiphoton interference on a boson sampler in the sparse regime},\ }\href {https://doi.org/https://doi.org/10.1103/PhysRevA.103.023722} {\bibfield  {journal} {\bibinfo  {journal} {Physical Review A}\ }\textbf {\bibinfo {volume} {103}},\ \bibinfo {pages} {023722} (\bibinfo {year} {2021})}\BibitemShut {NoStop}%
\bibitem [{\citenamefont {Ryser}(1963)}]{ryser1963combinatorial}%
  \BibitemOpen
  \bibfield  {author} {\bibinfo {author} {\bibfnamefont {H.~J.}\ \bibnamefont {Ryser}},\ }\href {https://doi.org/https://doi.org/10.5948/UPO9781614440147} {\emph {\bibinfo {title} {Combinatorial mathematics}}},\ Vol.~\bibinfo {volume} {14}\ (\bibinfo  {publisher} {American Mathematical Soc.},\ \bibinfo {year} {1963})\BibitemShut {NoStop}%
\bibitem [{Note7()}]{Note7}%
  \BibitemOpen
  \bibinfo {note} {For boosted fusion, the erasure probability is $p_{\protect \text {erasure}} = 1 - \protect \frac {7}{8}(1-p_{\protect \text {loss}})^4 = \protect \frac {1}{8} + \protect \frac {7}{2} p_{\protect \text {loss}} + \protect \mathcal {O}( p_{\protect \text {loss}}^2)$ \cite { bartolucci2023fusion}. Therefore, the additional photon loss introduced by photon distillation gates lead to an increased erasure probability $p_{\protect \text {erasure}} = \protect \mathcal {O}(1) \cdot p_{\protect \text {loss}}$.}\BibitemShut {Stop}%
\bibitem [{\citenamefont {de~Goede}\ \emph {et~al.}(2022)\citenamefont {de~Goede}, \citenamefont {Snijders}, \citenamefont {Venderbosch}, \citenamefont {Kassenberg}, \citenamefont {Kannan}, \citenamefont {Smith}, \citenamefont {Taballione}, \citenamefont {Epping}, \citenamefont {van~den Vlekkert},\ and\ \citenamefont {Renema}}]{de2022high}%
  \BibitemOpen
  \bibfield  {author} {\bibinfo {author} {\bibfnamefont {M.}~\bibnamefont {de~Goede}}, \bibinfo {author} {\bibfnamefont {H.}~\bibnamefont {Snijders}}, \bibinfo {author} {\bibfnamefont {P.}~\bibnamefont {Venderbosch}}, \bibinfo {author} {\bibfnamefont {B.}~\bibnamefont {Kassenberg}}, \bibinfo {author} {\bibfnamefont {N.}~\bibnamefont {Kannan}}, \bibinfo {author} {\bibfnamefont {D.}~\bibnamefont {Smith}}, \bibinfo {author} {\bibfnamefont {C.}~\bibnamefont {Taballione}}, \bibinfo {author} {\bibfnamefont {J.~P.}\ \bibnamefont {Epping}}, \bibinfo {author} {\bibfnamefont {H.}~\bibnamefont {van~den Vlekkert}},\ and\ \bibinfo {author} {\bibfnamefont {J.~J.}\ \bibnamefont {Renema}},\ }\bibfield  {title} {\bibinfo {title} {High fidelity 12-mode quantum photonic processor operating at ingaas quantum dot wavelength},\ }in\ \href@noop {} {\emph {\bibinfo {booktitle} {Integrated Photonics Research, Silicon and Nanophotonics}}}\ (\bibinfo {organization} {Optica Publishing Group},\ \bibinfo {year} {2022})\ pp.\ \bibinfo
  {pages} {ITu4B--3}\BibitemShut {NoStop}%
\end{thebibliography}%

\appendix
\section{Validity regime of first-order error approximation}
\label{appendix:firstordererror}
\begin{figure}[h!]
    \centering
    \includegraphics[width=8.6cm, height=20cm, keepaspectratio]{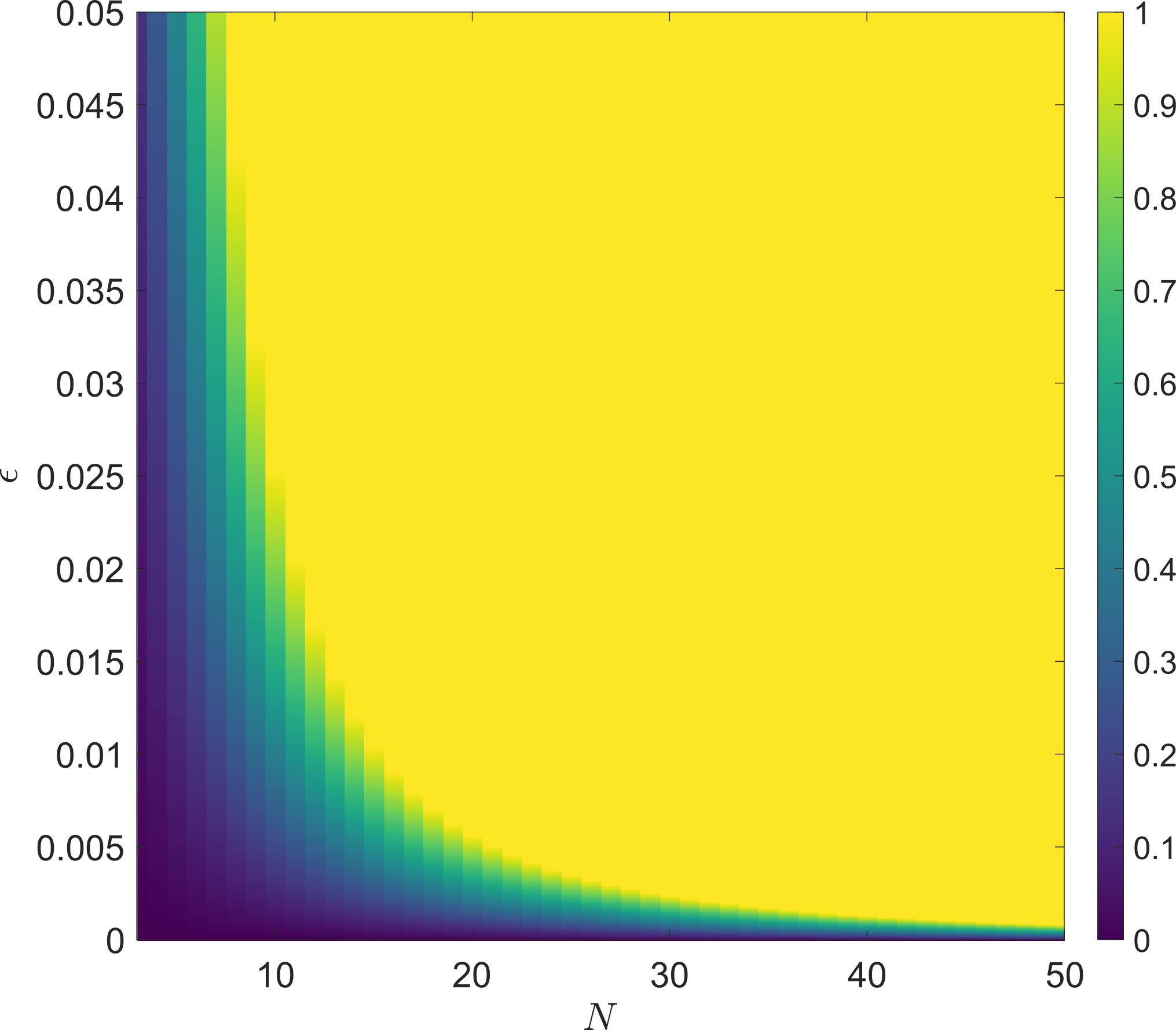}
    \caption{\textbf{Parametric scan of single-error regime}. The figure shows the ratio of multi-error probability $\delta_N(\epsilon)$ over partial distinguishability error $\epsilon$ is plotted. The low-error approximation is valid in the regime where $\frac{\delta_N(\epsilon)}{\epsilon} \ll 1$. Note that $\frac{\delta_N(\epsilon)}{\epsilon} \geq 1$ is set to 1 to improve readability.}
    \label{fig:errorapprox}
\end{figure}

Throughout out work, we work in the low-error regime, which we  define as the regime where Eq. \ref{eq:MultiMoylett} holds, i.e. where at most one photon is in an error mode. Here, we point out that the regime over which this approximation is valid shrinks with increasing number of photons.  
The probability of measuring a product state where $k$ out of $N$ photons are indistinguishable is given by
\begin{equation}
    P(k,N) = \binom{N}{k}(1-\epsilon)^k \epsilon^{N-k},
\end{equation}
where $1-\epsilon$ is the success probability as defined in Eq. \ref{eq:SingleMoylett}. It follows that the 0-error probability is given by $P(N,N) = (1-\epsilon)^N$ and the 1-error probability is given by $P(N-1,N) = N(1-\epsilon)^{N-1} \epsilon$. \\
\\
We define the probability of more-than-one error as 
\begin{equation}
    \delta_N(\epsilon) := \sum_{k=2}^{N} P(k,N). 
\end{equation}
Since the total probability is conserved, we find
\begin{equation}
    \delta_N(\epsilon) := 1 – (1-\epsilon)^N – N\epsilon(1-\epsilon)^{N-1}.
\end{equation}
To study photon distillation in the low error regime, we note that Eq. \ref{eq:MultiMoylett} is valid if the probability of having more than 1 error in a product state of $N$ imperfect photons is (much) smaller than the partial distinguishability error in one imperfect photon, i.e. $\delta_N(\epsilon) \ll \epsilon$ or $\frac{\delta_N(\epsilon)}{\epsilon} \ll 1$. Fig. \ref{fig:errorapprox} provides a parametric plot of the fraction $\frac{\delta_N(\epsilon)}{\epsilon}$ as a function of $N$ and $\epsilon$. We observe that the validity range of $\epsilon$ shrinks with increasing $N$. 

\section{Simulation methods}
\label{appendix:simulationmethods}
\begin{figure}[h!]
    \centering
    \includegraphics[width=8.6cm, height=20cm, keepaspectratio]{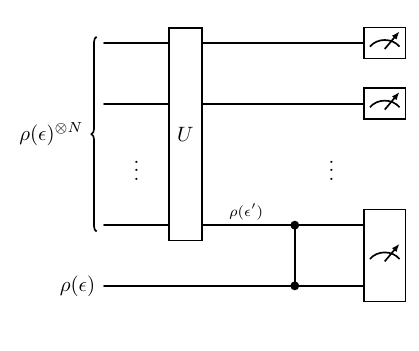}
    \caption{\textbf{Distillation simulation setup.} $N$ partial distinguishable photons are transformed by a $N$-mode Fourier transform $U$ (Eq. \ref{eq:Fourierdef}) into a multiphoton state. The partial state in the first $N-1$ auxiliary modes is measured. A successful measurement heralds the presence of a single photon state with reduced error $\epsilon^\prime < \epsilon$ (Eq. \ref{eq:SingleMoylett}) in mode $N$. The new error $\epsilon^\prime$ is extracted by first evaluating the Hong-Ou-Mandel visibility of a distilled photon and a non-distilled photon by interference in a 50:50 beamsplitter. One unitary transformation matrix simultaneously describes all transformations to simulate the interference of $N+1$ partially distinguishable photons.}
    \label{fig:fig2_setup}
\end{figure}

 Fig. \ref{fig:fig2_setup} shows the optical circuit which we use to simulate photon distillation. The product state $\rho(\epsilon)^{\otimes N}$ is transformed by a $N$-mode Fourier transform into a multiphoton entangled state. A given partial measurement of the multiphoton output state in the first $N-1$ modes heralds the presence of a single photon state $\rho(\epsilon^\prime)$ in mode $N$. Subsequently, the product state $\rho(\epsilon^\prime) \otimes \rho(\epsilon)$ is interfered by a balanced beam splitter. We extract the reduced partial distinguishability error $\epsilon^\prime$ by computing first the new visibility  $V^\prime = \text{Tr}[\rho(\epsilon)\rho(\epsilon^\prime)] = (1-\epsilon)(1-\epsilon^\prime)$ for known $\epsilon$. We compute the counting statistics behind the balanced beam splitter for a given herald measurement by using the formalism for interference of partially distinguishable photons as presented in Ref.~\cite{tichy2015sampling}. Our calculations assume perfectly functioning and loss-free components. Although studying the interference of two equally purified photons seems simpler, this actually requires the computation of matrix permanents of size $2N$x$2N$. Our setup only requires matrix permanent computations of size $(N+1)$x$(N+1)$, which is beneficial because these computations are intrinsically inefficient \cite{ryser1963combinatorial}. Furthermore, our setup can be experimentally implemented to study photon distillation schemes with a less resource-intensive approach, as it uses only one distilled photon instead of two \cite{faurby2024purifying}. 

\section{Multi-photon interference coefficients}
\label{app:conjecture}
The probability of output photon distribution $i$ can be written in the formalism of Ref. \cite{tichy2015sampling, renema2018efficient} as 
\begin{equation}
    q(\epsilon) := \sum_{j=0}^{N} c_j (1-\epsilon)^j,
\end{equation}
where the $j$-th order multiphoton interference coefficients are defined by:
\begin{equation}
    c_j := \frac{1}{\mu_i} \sum_{\sigma^j } \text{perm}[M \circ \bar{M}_{\sigma^j}].
\end{equation}
Here, $\mu_i$ is a multiplicity correction $\mu(s) = \prod_i (s_i)!$ for outputs with more than one photon in the same mode, $M$ is an $N$-by-$N$ submatrix of unitary transformation matrix $U$ (where rows and columns of $M$ correspond to input and output modes, respectively), $\sigma^j$ corresponds to all row permutations $\sigma$ with $N-j$ fixed-point elements, $\circ$ denotes the element-wise matrix product and $\bar{M}_{\sigma^j}$ is the complex conjugate of the matrix $M$ with rows permuted according to $\sigma^j$ (for detail see \cite{renema2018efficient}). \\

\begin{figure}[h!]
    \centering
    \includegraphics[width=8.6cm, height=20cm, keepaspectratio,]{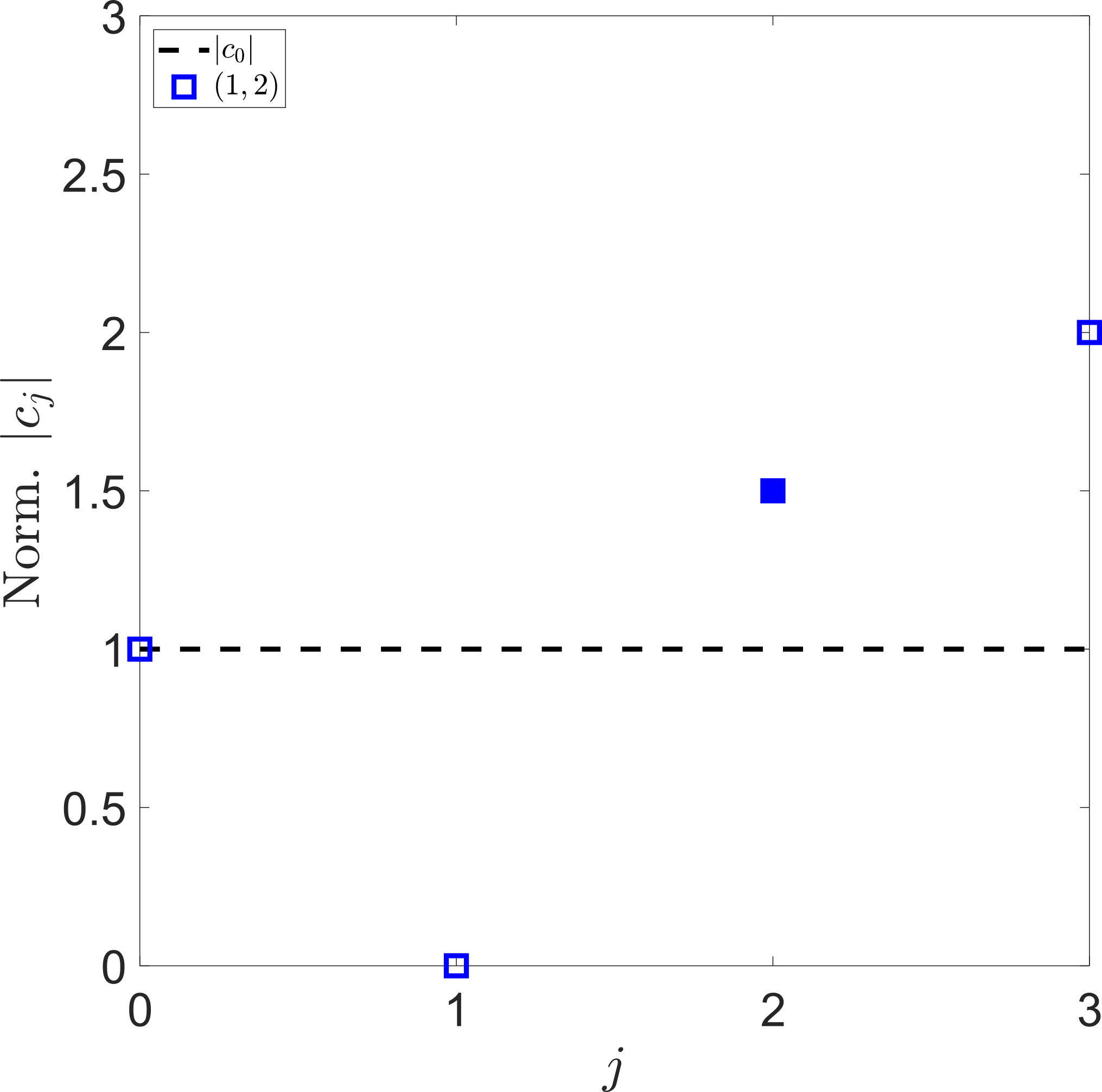}
    \caption{\textbf{Multiphoton interference coefficients for the $N = 3$ Fourier transform-based distillation scheme.}. Absolute values of the $j$-th order multiphoton interference contribution coefficients $c_j$ are plotted for herald measurement $(1,2)$. Open markers indicate a positive contribution while filled markers indicate a negative contribution. All coefficients are normalized with respect to $c_0$, which is the probability of observing that detection event in case of fully distinguishable photons.}
    \label{fig:c_j_N_3_Fourier}
\end{figure}

First, we analyze the $c_j$ distribution of the $N = 3$ photon distillation scheme proposed by Ref. \cite{marshall2022distillation}. We focus specifically on herald measurement list $(1,2)$, which corresponds to output photon distribution $(1,2,3)$. The (normalized) coefficient distribution is presented in Fig. \ref{fig:c_j_N_3_Fourier}. We observe that the (absolute value of) 2- and 3-photon interference contributions are relatively high compared to the zero photon interference contribution $c_0$, which corresponds to the transmission probability for fully distinguishable photons. By definition, $c_1 = 0$. One can verify $\sum_{j=0}^{3}c_j j = 3c_0$ and $2\sum_{j=0}^{3}c_j = 3c_0$, so the $N = 3$ Fourier transform-based distillation scheme satisfies Eq. \ref{eq:conjecture}.\\

\begin{figure}[h]
    \centering
    \includegraphics[width=8.6cm, height=20cm, keepaspectratio,]{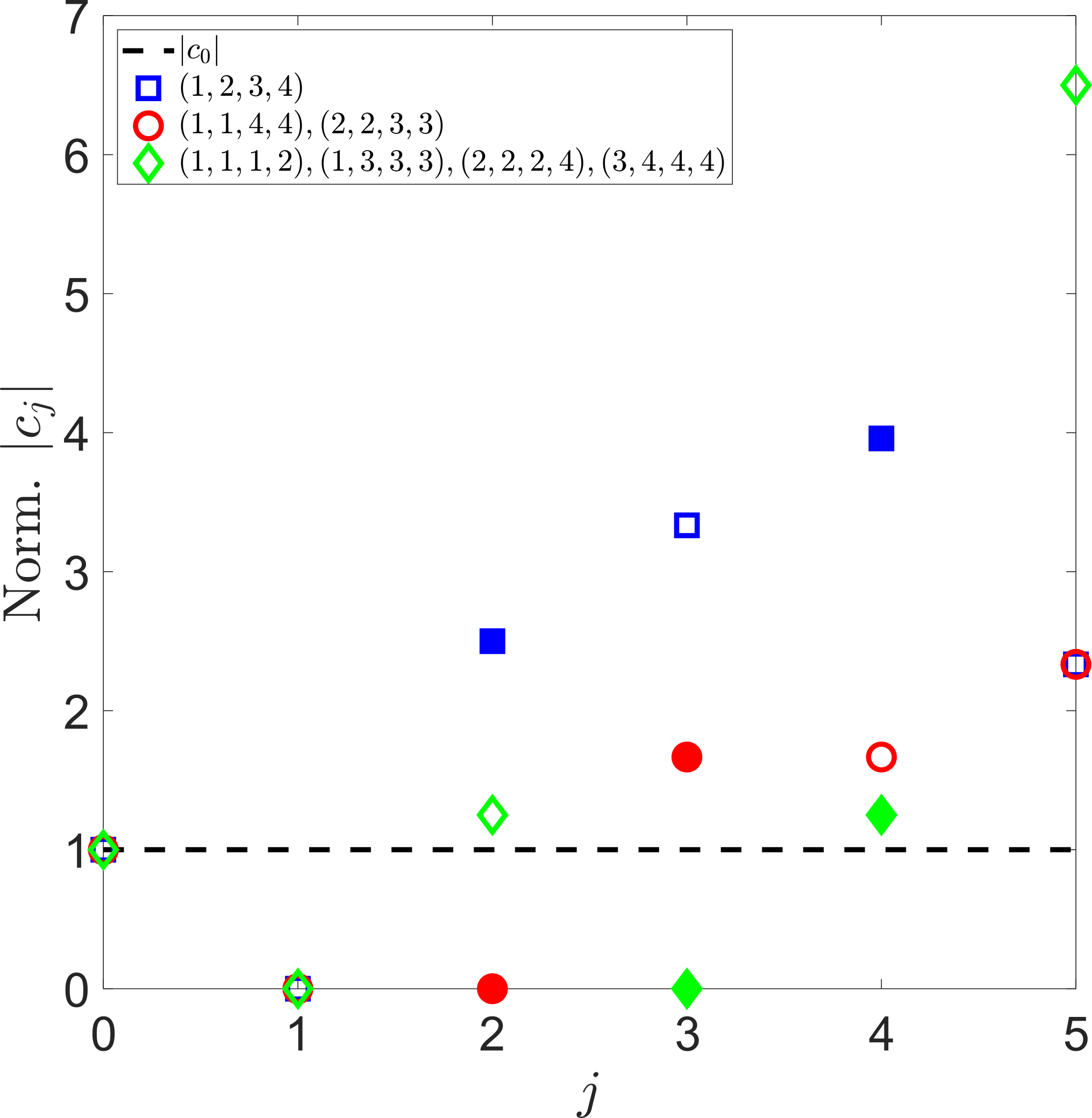}
    \caption{\textbf{Multiphoton interference coefficients for the $N = 5$ Fourier transform-based distillation scheme.}. Absolute values of the $j$-th order multiphoton interference contribution coefficients $c_j$ are plotted for herald measurements $l_i = (m_1, m_2, m_3, m_4)$ that specify how the remaining 4 photons are distributed among the first 4 modes, while the distilled photon is in mode 5. Open markers indicate a positive contribution while filled markers indicate a negative contribution. All coefficients are normalized with respect to $c_0$, which is the probability of observing that detection event in case of fully distinguishable photons.}
    \label{fig:c_j_N_5_Fourier}
\end{figure}

Now we turn our attention to the $c_j$ distribution for the $N = 5$ photon Fourier transform-based distillation scheme discussed in Sec. \ref{sec:FTBP}, which is presented in Fig. \ref{fig:c_j_N_5_Fourier}. Similar to Fig. \ref{fig:c_j_N_3_Fourier}, the distribution for $N = 5$ shows for all individual herald measurements the remarkable feature $\sum_{j=0}^{5}c_j j = 4\sum_{j=0}^{5}c_j$ satisfying again Eq. \ref{eq:conjecture}. For reference, we have $\sum_{j=0}^{5}c_j = \frac{5}{24}c_0$ (blue squares), $\frac{10}{3}c_0$ (red circles), $\frac{15}{2}c_0$ (green diamonds). Motivated by inspection of Fig. \ref{fig:fig3_resulterror} and \ref{fig:resultheraldprob}, we suspect that larger ratios $\sum_{j=0}^{N}c_j$ over $c_0$ indicate better performing photon distillation herald measurements. \\

\begin{figure}[h!]
    \centering
    \includegraphics[width=8.6cm, height=20cm, keepaspectratio,]{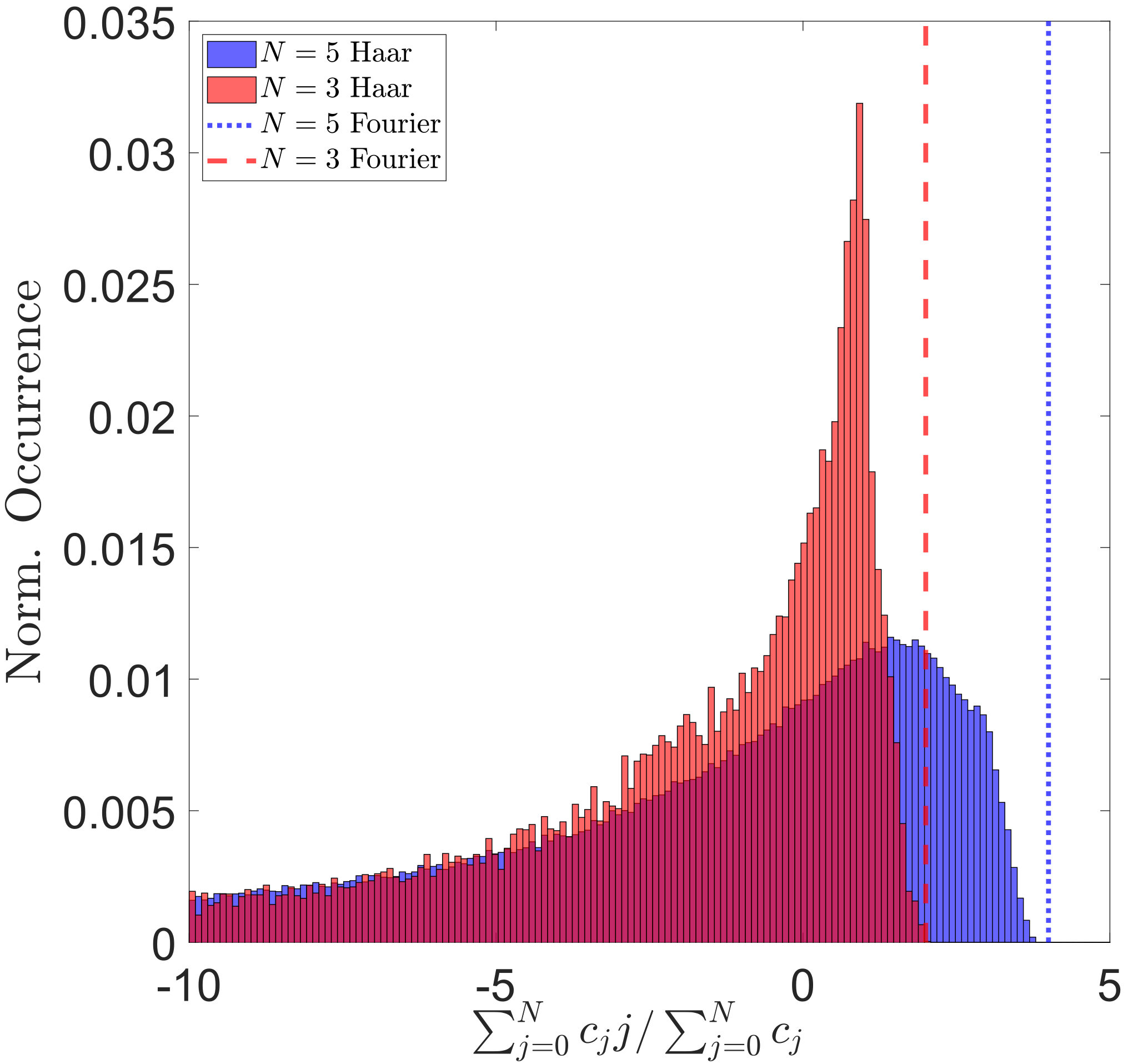}
    \caption{\textbf{Multiphoton interference coefficients for Haar-random unitary transformations.} The normalized occurrence of the ratio $(\sum_{j=0}^{N}c_j j) / (\sum_{j=0}^{N}c_j)$ for all mode-assignment lists with exactly one photon in mode $N$ for $10^4$ random Haar matrices is presented for $N = 3,5$. For comparison, the ratios calculated for relevant Fourier transform-based herald measurements are indicated by dashed/dotted lines.}
    \label{fig:c_j_Haar}
\end{figure}

For comparison, we inspected the $c_j$ distribution for Haar-random unitary matrices. We generate $10^4$ Haar-random matrices and compute the ratio $\sum_{j=0}^{N}c_j j$ over $\sum_{j=0}^{N}c_j$ for all mode-assignment lists with exactly one in mode $N$. The histogram of this ratio, for $N = 3$ and $N = 5$ is plotted in Fig. \ref{fig:c_j_Haar}. We observe that the herald measurements for the Fourier transforms (dashed and dotted lines) have extreme ratios $N-1$ that satisfy Eq. \ref{eq:conjecture}. These observed extreme ratios are close to the maximum allowed ratio. Since the equality Eq. \ref{eq:TwoStep} = Eq. \ref{eq:FromModel} must hold, we require 
\begin{equation}
    \label{eq:equalityRatioderivation}
    \sum_{j=0}^{N} c_j - \epsilon \sum_{j=0}^{N}c_j j = \sum_{j=0}^{N} c_j (1-N \epsilon) + \lambda N \epsilon,
\end{equation}
where probability $\lambda \geq 0$ to remain physically relevant. Eq. \ref{eq:equalityRatioderivation} is simplified to find
\begin{equation}
    \label{eq:lambda_explicit}
    \lambda = \sum_{j=0}^{N} c_j - \left( \sum_{j=0}^{N}c_j j \right) /N.
\end{equation}
Finally, $\lambda \geq 0$ gives
\begin{equation}
    \label{eq:maxbound}
    N \geq  \frac{\sum_{j=0}^{N}c_j j}{\sum_{j=0}^{N} c_j}.
\end{equation}
Given that $N \geq N-1$, we conclude that Fourier transforms are optimal in the asymptotically large $N$ regime.  
\\

\begin{table}[b]
\caption{\label{tab:summary_numerical_study}
\textbf{Summary of numerical analysis} for $N$-photon Fourier transform. Number of allowed outcomes where exactly one photon ends in mode $N$ ($N_\text{allowed}$), maximum calculated numerical difference based on Eq. \ref{eq:conjecture_difference} ($\Delta_\text{max}$), number of allowed outcomes that constitute a photon distillation strategy ($N_\text{herald}$), and percentage of allowed outcomes that constitute a photon distillation strategy.
}
\begin{ruledtabular}
\begin{tabular}{lllll}
\textrm{$N$}&
\textrm{$N_\text{allowed}$}&
\textrm{$\Delta_\text{max}$}&
\textrm{$N_\text{herald}$}&
\textrm{$N_\text{herald} / N_\text{allowed}$ ($\%$)}\\
\colrule
3 & 1 & 0 & 1 & 100\\
4 & 2 &$6\cdot 10^{-17}$& 2 & 100\\
5 & 7 & $6\cdot 10^{-17}$& 7 & 100 \\
6 & 20 & $6\cdot 10^{-17}$ & 14 & 70 \\
7 & 66  & $8\cdot 10^{-16}$ & 66 & 100\\
8 & 212 & $4\cdot 10^{-17}$ & 212 & 100 \\
9 & 715 & $6\cdot 10^{-16}$ & 715 & 100 \\
10 & 2424 & $2\cdot 10^{-15}$ & 1944 & 80 \\
\end{tabular}
\end{ruledtabular}
\end{table}

We summarize the results of our numerical studies for $N = 3$ through $N = 10$ in Tab. \ref{tab:summary_numerical_study}. For each allowed outcome with exactly one photon in mode $N$, we calculated the multiphoton coefficients. For each of these allowed outcomes, we checked our conjecture (Eq. \ref{eq:conjecture}) by calculating the difference 
\begin{equation}
    \Delta := \left|\sum_{j=0}^{N}c_j j - (N-1) \sum_{j=0}^{N}c_j \right|,
    \label{eq:conjecture_difference}
\end{equation}
which is ideally zero and which is a measure of the numerical accuracy of our calculations. Then, we calculated the number of allowed outcomes that are non-suppressed (and therefore relevant to photon distillation). In the case of $N = 6$ and $N = 10$ we find that some allowed outcomes have probability 0, which means they do not result in photon distillation, consistent with the  results of Ref. \cite{tichy2010zero}.\\

Finally, we prove Eq. \ref{eq:conjecture} for single-mode bunched outcomes. First, we combine Eq. \ref{eq:TwoStep} and \ref{eq:FromModel} to find $\sum_{j=0}^{N}c_j j = (q_\text{i} - \lambda)N$. Second, we note that the single-mode bunching probability  $p_\text{single} = \text{perm}[S] c_0$ is determined by the indistinguishability matrix $S$ \cite{seron2023boson}. For fully indistinguishable photons, $\text{perm}[S]=N!$ which results in $q_\text{i}=N! c_0$. However, if there is one fully distinguishable photon, $\text{perm}[S]=(N-1)!$ because there exists a row and a column in $S$ for which all the elements, except for the diagonal, are zero. Therefore, we find $\lambda=(N-1)! c_0$. Combining the above results, we derive $\sum_{j=0}^{N}c_j j = (N-1)N!c_0 = (N-1)\sum_{j=0}^{N} c_j $, which completes our proof.

\section{Herald probability}
\label{app:heraldprob}
\begin{figure}[h!]
    \centering
    \includegraphics[width=8.6cm, height=20cm, keepaspectratio]{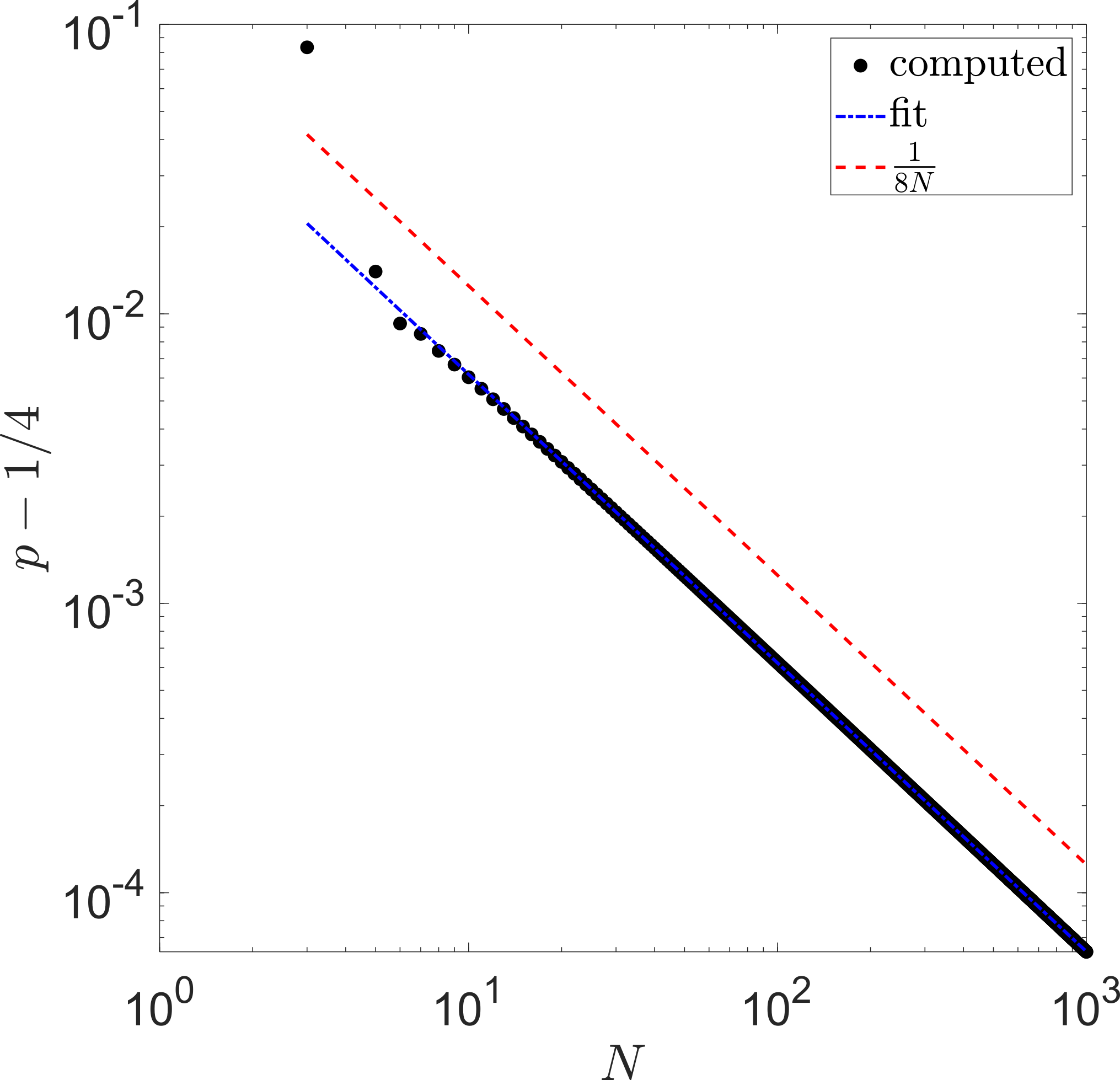}
    \caption{\textbf{Herald probability for $N$-photon Fourier transform-based distillation schemes}. Excluding $N=3,4$, we fit $p - 1/4 = 0.0614\cdot N^{-0.9973}$.}
    \label{fig:p_numeric}
\end{figure}

To compute the total herald probability $p$, we focus on the marginal probability of observing exactly one photon in a particular mode of interest. Without loss of generality, we assume that the distilled photon is located in output mode 1. Using the expanded sample space formalism introduced by Clifford \& Clifford \cite{clifford2018classical}, we focus first on the marginal probability where the first $k$ out of $N$ photons end up in output mode 1: 
\begin{equation}
	p^{(k)} := p(1_1, \dots, 1_k).
	\label{eq:C&Cpk}
\end{equation}
According to Lemma 1 in Ref. \cite{clifford2018classical}, we find that the marginal probability of the subsequence $(1_1, \dots 1_k)$ is given by
\begin{equation}
	p^{(k)} = \frac{1}{k!} |\text{perm}{M}|^2,
\end{equation}
where we used that, in case of Fourier transforms, all $\binom{N}{k}$ input - output combinations share the same submatrix $M$. Here, submatrix $M$ is a rank 1 matrix of size $k$ x $k$ where the matrix elements are given by $M_{ij} = \frac{1}{\sqrt{N}}$ such that $\text{perm}{(M)} = \frac{k!}{\sqrt{N}^k}$. Therefore, we have 
\begin{equation}
	p^{(k)} = \frac{k!}{N^k}.
	\label{eq:p(k)}
\end{equation}
However, the marginal probability $p^{(k)}$ still includes the possibility that any of the $N-k$ leftover photons end in mode 1. We introduce the new marginal probability $\tilde{p}^{(k)}$ where none of the remaining $N-k$ photons end in mode 1, by definition. We can expand $p^{(k)}$ in terms of $\tilde{p}^{(k+j)}$ for $j \in \{0, 1, \dots, N-k \}$ such that the first $k + j$ photons are in output mode 1 while the leftover $N - k - j$ photons are excluded. For each $j$, there are $\binom{N-k}{j}$ equivalent probabilities. Therefore, we find
\begin{equation}
    p^{(k)} = \sum_{j = 0}^{N-k} \binom{N-k}{j}\tilde{p}^{(k+j)}.
    \label{eq:expansionCC}
\end{equation}
Inversion of Eq. \ref{eq:expansionCC} results in
\begin{equation}
    \tilde{p}^{(k)} = \sum_{j = 0}^{N-k} (-1)^j \binom{N-k}{j} p^{(k+j)}.
    \label{eq:ptilde}
\end{equation}
For photon distillation purposes considered in this paper, we are interested in the particular case where $k = 1$. Substituting Eq. \ref{eq:p(k)} in Eq. \ref{eq:ptilde}, we find
\begin{equation}
    \tilde{p}^{(1)} = \frac{1}{N} \sum_{j=0}^{N-1} (-1)^j(j+1) \binom{N-1}{j} \frac{j!}{N^j}. 
\end{equation}
The last terms can be rewritten as
\begin{equation}
    \binom{N-1}{j} j! = N^j \prod_{i=1}^{j}\left(1-\frac{i}{N}\right),
\end{equation}
such that
\begin{equation}
    \tilde{p}^{(1)} = \frac{1}{N} \sum_{j=0}^{N-1} (-1)^j(j+1) \prod_{i=1}^{j}\left(1-\frac{i}{N}\right). 
\end{equation}
As there are $\binom{N}{1} = N$ equivalent probabilities, we have $p = N\tilde{p}^{(1)}$. Therefore, we arrive at Eq. \ref{eq:p}: 
\begin{equation}
    p = \sum_{j=0}^{N-1} (-1)^j (j+1) \prod_{i = 1}^{j} \left(1-\frac{i}{N}\right).
\end{equation}
We evaluated $p$ numerically for $N = 3 - 1000$. In Fig. \ref{fig:p_numeric}, we observe that $p = \frac{1}{4} + \mathcal{O}(\frac{1}{N})$. Therefore, our computations suggest that $p$ tends to converge to $\frac{1}{4}$ after initial finite-size effects at $N = 3,4$.

\section{Photon loss}
\label{app:freelunch}
In realistic photon distillation gates, photon loss errors reduce the gate success probability and increase the erasure error. First, the photon distillation gate fails automatically if more than one photon is lost, as $N-1$ photons must be measured in the herald outputs for the gate to succeed. Therefore, higher-order loss errors result in a lower herald probability. However, a lower herald probability can be mitigated through multiplexing, thus higher-order photon loss errors are locally confined. Secondly, the photon distillation gate can still yield a valid herald measurement if only one photon is lost, which projects the distilled output state onto the vacuum state instead of a single-photon state with reduced indistinguishability error. Therefore, first-order photon loss errors result in an erasure error, but they do \textit{not} reduce the indistinguishability error reduction performance. Unfortunately, this erasure error is unheralded, thus first-order photon loss errors are not bound to the distillation gate. As these erasure errors can propagate through entangling operations, they can potentially corrupt a logical qubit and are therefore an unwanted side effect of photon distillation gates. \\

As discussed in the main text, indistinguishability errors result in measurement errors, and a reasonable approximation for the measurement error is $p_{\text{error}} = \mathcal{O}(1) \cdot \epsilon$. Similarly, photon loss errors result in erasure errors, and the first-order approximation for the erasure error is $p_{\text{erasure}} = \mathcal{O}(1) \cdot p_{\text{loss}}$ \footnote{For boosted fusion, the erasure probability is $p_{\text{erasure}} = 1 - \frac{7}{8}(1-p_{\text{loss}})^4 = \frac{1}{8} + \frac{7}{2} p_{\text{loss}} + \mathcal{O}( p_{\text{loss}}^2)$ \cite{ bartolucci2023fusion}. Therefore, the additional photon loss introduced by photon distillation gates lead to an increased erasure probability $p_{\text{erasure}} = \mathcal{O}(1) \cdot p_{\text{loss}}$.}. It is evident that photon distillation reduces $p_{\text{error}}$ while simultaneously increasing $p_{\text{erasure}}$. However, the threshold for measurement errors in absence of other error types is one order of magnitude lower than those for erasure errors, i.e. $p_{\text{erasure,th}} \approx 10\cdot p_{\text{error,th}}$ \cite{bartolucci2023fusion}. Therefore, there is a large area in parameter space where the combination of measurement errors and erasure errors enabled by the photon distillation protocol is still beneficial even in the light of the erasure errors introduced by the same protocol.\\

For our preliminary analysis, we consider similar resources as used in Ref. \cite{faurby2024purifying}. We estimate an input indistinguishability error $\epsilon = 0.0204$, based on the reported quantum dot optimal $V_0 = 0.9050$ and $g^{(2)}(0) = 0.0279$ via $(1-\epsilon)^2 = \frac{V_0 + g^{(2)}(0)}{1 - g^{(2)}(0)}$ \cite{ollivier2021hong}. Realistic values for the erasure error are based on the reported losses for state-of-the-art low-loss quantum photonic processors that are compatible with the wavelength of aforementioned quantum dots \cite{de2022high}. Here, a 3.4 dB insertion loss is reported for a processor that implements 12-mode transformations. With fixed coupling losses of 1.8 dB  \cite{taballione202320}, this results in a $N$-mode photon distillation gate insertion loss of $IL = 1.8 + \frac{1.6}{12} N$ [dB]. The added photon distillation gate loss, and hence the increase in erasure error, is found by $p_{\text{loss}} = 1 – 10^{-\frac{IL}{10}}$. \\

\begin{figure}[h!]
    \centering
    \includegraphics[width=8.6cm, height=20cm, keepaspectratio]{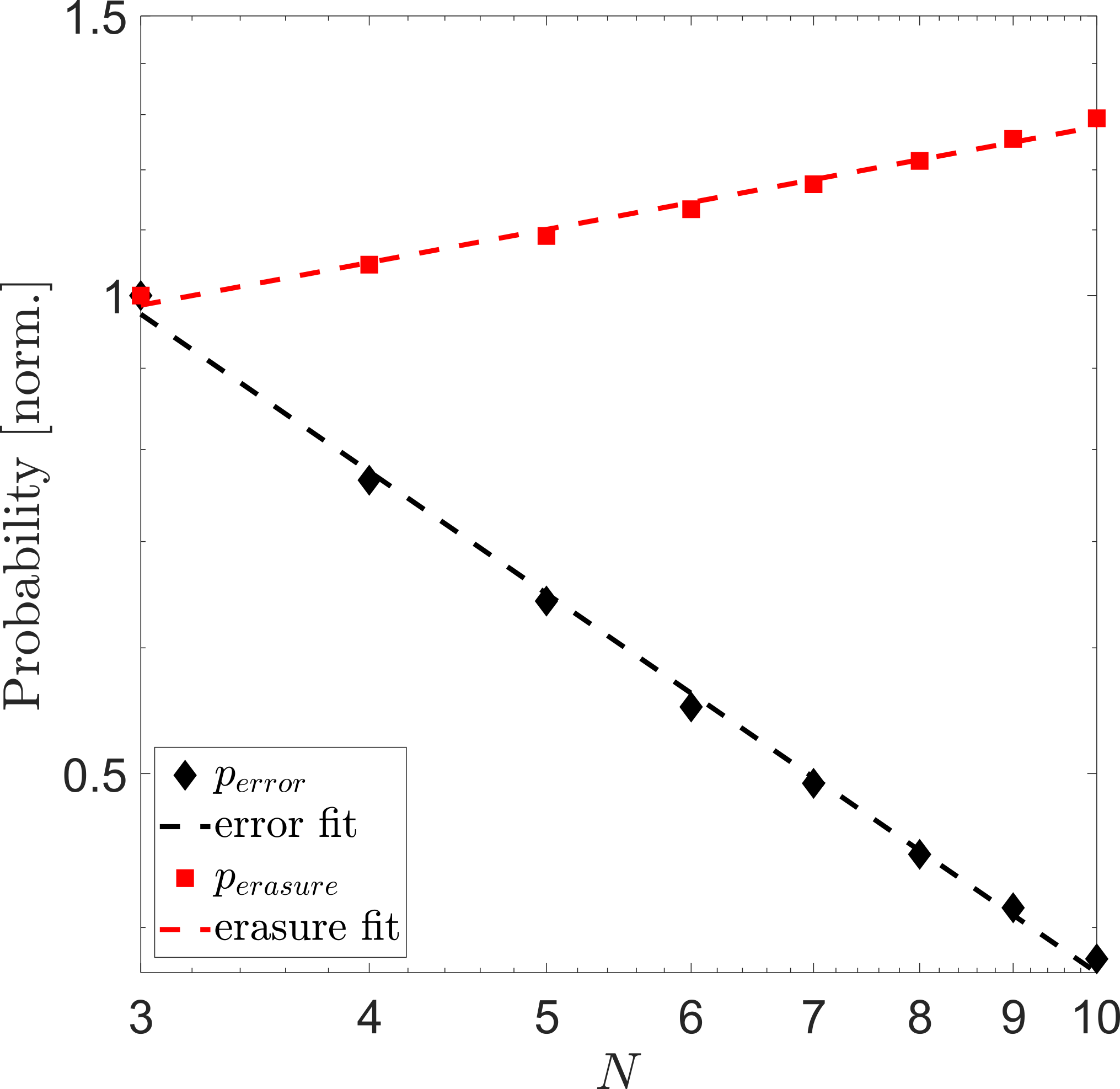}
    \caption{\textbf{Trade-off for realistic $N$-photon Fourier transform-based distillation schemes}. Both measurement and erasure errors are normalized relative to $N = 3$. We heuristically fit $p_{\text{error}} \propto N^{-0.7930}$ and $p_{\text{erasure}} \propto N^{0.2150}$.}
    \label{fig:error_erasure}
\end{figure}

To emphasize the trade-off between realistic photon distillation schemes, we normalize the calculated measurement and erasure errors to the results of the $N = 3$ scheme, which corresponds to the original scheme of Ref. \cite{marshall2022distillation}. By normalizing relative to the state-of-the-art $N = 3$ scheme, we can clearly visualize the trade-off for realistic photon distillation schemes. In Fig. \ref{fig:error_erasure}, we observe that the relative decrease in measurement error is larger than the inevitable increase in erasure error (at least up to $N = 10$). Since existing photonic quantum computer architectures are generally more robust to erasure errors than to measurement errors \cite{bartolucci2023fusion} (after all, erasure errors fall outside the code space, while measurement errors typically fall within the code space), we conclude that the proposed photon distillation gates do not constitute a 'cure' that is worse than the 'disease'. 

\end{document}